\documentclass[aip,graphicx,amsmath,amssymb,preprint]{revtex4-1}
\usepackage{CJKutf8}
\usepackage{lineno,hyperref}
\usepackage[normalem]{ulem}
\usepackage{graphics}
\usepackage{amsmath,bm}
\usepackage{amssymb}
\usepackage{graphicx}
\usepackage{bm}
\usepackage{amsfonts}
\usepackage{color}
\usepackage{epstopdf}
\usepackage{booktabs}
\usepackage{array}
\usepackage{hyperref}
\usepackage{graphics}
\usepackage{epsfig}
\usepackage{bm}
\usepackage{epic,eepic}
\usepackage{epsfig}
\usepackage{pifont}

\usepackage{subfigure}

\usepackage{xspace}
\usepackage{graphicx,color}
\usepackage{amsmath,amssymb,amsfonts,mathrsfs}
\usepackage{url}
\usepackage[normalem]{ulem}
\usepackage{booktabs}
\usepackage{float}

\usepackage[toc,page]{appendix}
\usepackage[capitalize]{cleveref}
\usepackage{mathtools}
\usepackage[normalem]{ulem}
\usepackage{gensymb}
\usepackage{algorithm}
\usepackage{algorithmic}
\usepackage{lineno}

\begin{document}

\title{Physics informed data-driven near-wall modelling for lattice Boltzmann simulation of high Reynolds number turbulent flows}






\begin{CJK*}{UTF8}{gbsn}
\author{Xiao Xue}

\affiliation{ 
Division of Fluid Dynamics, Department of Mechanics and Maritime Sciences, Chalmers University of Technology, 41296, Gothenburg, Sweden
}
\affiliation{ 
Centre for Computational Science, Department of Chemistry, University College London, London, UK
}
\author{Shuo Wang}
\affiliation{ 
Fluids \& Flows Group, Department of Applied Physics, Eindhoven University of Technology, The Netherlands
}
\author{Hua-Dong Yao}
\affiliation{ 
Division of Fluid Dynamics, Department of Mechanics and Maritime Sciences, Chalmers University of Technology, 41296, Gothenburg, Sweden
}
\author{Lars Davidson}%
\affiliation{ 
Division of Fluid Dynamics, Department of Mechanics and Maritime Sciences, Chalmers University of Technology, 41296, Gothenburg, Sweden
}
\author{Peter V. Coveney}%
\email{p.v.coveney@ucl.ac.uk}
\affiliation{ 
Centre for Computational Science, Department of Chemistry, University College London, London, UK
}
\affiliation{ 
Informatics Institute, University of Amsterdam, The Netherlands
}
\affiliation{
Centre for Advanced Research Computing, University College London, UK
}

\date{\today}

\begin{abstract}
Data-driven approaches offer novel opportunities for improving the performance of turbulent flow simulations, which are critical to wide-ranging applications from wind farms and aerodynamic designs to weather and climate forecasting. While conventional continuum Navier-Stokes solvers have been the subject of a significant amount of work in this domain, there has hitherto been very limited effort in the same direction for the more scalable and highly performant lattice Boltzmann method (LBM), even though it has been successfully applied to a variety of turbulent flow simulations using large-eddy simulation (LES) techniques. In this work, we establish a data-driven framework for the LES-based lattice Boltzmann simulation of near-wall turbulent flow fields. We do this by training neural networks using improved delayed detached eddy simulation data. Crucially, this is done in combination with physics-based information that substantially constrains the data-driven predictions.  Using data from turbulent channel flow at a friction Reynolds number at $5200$, our simulations accurately predict the behaviour of the wall model at arbitrary friction Reynolds numbers up to $1.0 \times 10^6$.  In contradistinction with other models that use direct numerical simulation datasets, our physics-informed model requires data from very limited regions within the wall-bounded plane, reducing by three orders of magnitude the quantity of data needed for training. We also demonstrate that our model can handle data configurations when the near-wall grid is sparse. Our physics-informed neural network approach opens up the possibility of employing LBM in combination with highly specific and therefore much more limited quantities of macroscopic data, substantially facilitating the investigation of a wide-range of turbulent flow applications at very large scale.

\end{abstract}

\keywords{Physics-informed neural networks, wall model, lattice Boltzmann method, large-eddy-simulation}

\maketitle
\end{CJK*}
\section{INTRODUCTION}
Wall-modelled large-eddy simulations (WMLES) are crucial in within industrial design aerodynamics-driven applications such as aircraft, high-speed trains and wind turbines~\cite{porte2011large, mehta2014large}. These simulations provide a balance between computational efficiency and the resolution of flow physics. Although large-eddy simulations (LES) require fewer computational grid points than direct numerical simulations (DNS), wall-resolved LES necessitates fine grid resolution near the wall. The grid point requirement scales with $O(Re^{13/7})$ \cite{chapman1979computational, choi2012grid, yang2021grid}. To expedite the aerodynamics-driven design process, WMLES emerges as a viable alternative to reduce computational costs significantly. Accurate modelling of the near-wall fluid field is imperative to maintain coarse-grained near-wall field resolution without compromising the accuracy of far-field physics. Traditionally, the near-wall field is modelled using the law of wall or by solving thin-boundary-layer equations\cite{schumann1975subgrid, park2014improved, larsson2016large, bose2018wall}. In the current era of big data, the increasing availability of DNS data promotes machine learning and data-driven approaches to accurately capture the physics of coarse-grained models~\cite{brunton2020machine}. 

Data-driven approaches based on machine learning (ML) have been widely used in industrial processes\cite{yin2014review}, computer vision\cite{gopalakrishnan2017deep} and  control related problems\cite{ding2014data, hou2013model}. Recently, the data-driven approach has been employed for subgrid scale (SGS) modelling\cite{sarghini2003neural, gamahara2017searching, wu2018physics, xie2019artificial, cai2021flow} and near wall modelling\cite{bae2022scientific, yang2019predictive}. A common limitation of data-driven approaches is their lack of generality which stems from an inadequate understanding of the underlying physics. This limitation is evident when trained ML-based models show excellent ability to predict scenarios they have been trained on, sometimes referred to as ``seen scenarios" or ``interpolated schemes", yet, struggle to predict ``unseen scenarios" or ``extrapolated schemes", which fall outside the scope of their training data.
A physics-informed data-driven approach offers a way of more effectively utilising datasets by embedding physical principles within the training process. Numerous studies have successfully implemented Physics-Informed Neural Networks (PINNs) for investigating turbulent flows using DNS data~\cite{raissi2019physics, wang2017physics, davidson2022using, yang2019predictive}. While effective, these methods rely on DNS data which is not only computationally demanding to produce but also requires substantial storage capacity. In contrast, Bae and Koumoutsakos\cite{bae2022scientific} employed multi-agent reinforcement learning to model the near-wall field in turbulent flows using hybrid Reynolds-Averaged Navier-Stokes (RANS)/LES data. The process of generating training data engenders a notable reduction in both computational and storage demands, particularly when contrasted with techniques that depend on DNS data. Such methods are invariably based on the Computational Fluid Dynamics (CFD) approaches to solve the Navier-Stokes equations. The Lattice Boltzmann Method (LBM) by contrast, is recognized for its ability to model complex fluid phenomena across multiple scales, from microscopic \cite{xue2020brownian, xue2018effects, xue2021lattice, chiappini2018ligament, chiappini2019hydrodynamic} to macroscopic \cite{hou1994lattice, toschi2009lagrangian, karlin1999perfect, lallemand2000theory}. Instead of solving the Navier-Stokes (NS) equations, LBM offers an alternative to conventional CFD methods by solving the Boltzmann equation at a mesoscopic level~\cite{succi2001lattice, lallemand2000theory, kruger2017lattice, lallemand2021lattice}. LBM is favoured for its parallelization-friendly characteristics, primarily due to the localized updating of discrete single particle distribution functions. Through the application of a Chapman-Enskog analysis to the Boltzmann equation, it is possible to derive the Navier-Stokes equations. Hou et al.\cite{hou1994lattice} were instrumental in integrating the LES-based approach with LBM, particularly in modelling effective turbulent viscosity. As for near-wall modelling in LBM, Malaspinas et al.\cite{malaspinas2014wall} demonstrated the successful reconstruction of first-layer near-wall velocity with a regularized scheme~\cite{latt2008straight} using the Musker wall function or logarithmic law~\cite{musker1979explicit}. Subsequent research focused on reconstructing the velocity field or modelling velocity bounce-back~\cite{malaspinas2014wall, haussmann2019large, maeyama2020unsteady, wilhelm2021new}. The forcing-based method, noted for its simplicity, has also been applied in modelling the near-wall fluid field~\cite{kuwata2021wall, xue2023wall}. \\
To date, existing data-driven wall models have been developed within the Navier-Stokes (NS) framework. However, there is no established data-driven approach that models the near-wall fluid field using the LBM framework. This paper presents a methodology for utilizing both grid-intensive and grid-sparse improved delayed detached eddy simulation (IDDES)~\cite{shur2008hybrid} data to develop a generalized neural network(NN) wall model within the lattice Boltzmann framework. In the current study, IDDES data is trained at friction Reynolds number, $Re_{\tau}$, equal to $Re_{\tau} = 5200$. Subsequently, this data is applied to the NN wall model for turbulent channel flow, incorporating a synthetic turbulent generator (STG) at the inlet as described by Xue et al. \cite{xue2022synthetic}, at $Re_{\tau} = 1000, 2000, 5200$. The results show good agreement with DNS data at coarse resolutions with the channel height of 20 lattice grid separations. To highlight the model's generality, we further validate our NN wallmodel for the LBM based turbulent channel flow at the friction Reynolds number $Re_{\tau} = 1.0 \times 10^5$ and $Re_{\tau} = 1.0 \times 10^6$. It is noteworthy that the grid-intensive data used in this study amounts to only 45 megabytes (MBs) while the grid-sparse data is even more compact, requiring just 6.4 MBs, in contrast to DNS data, which can range from hundreds of gigabytes (GBs) to several terabytes (TBs). Our work illustrates the benefits of integrating data-driven approaches with physical knowledge to model the near-wall dynamics in the lattice Boltzmann method. We demonstrate here that this approach can leverage CFD data very effectively in the context of NN-based wall modelling within the lattice Boltzmann framework. 

\section{Results}\label{sec:results}
\subsection{Physics-informed neural network wall modelling framework for LBM}
\begin{figure}
\centering
\includegraphics[width=1\textwidth]{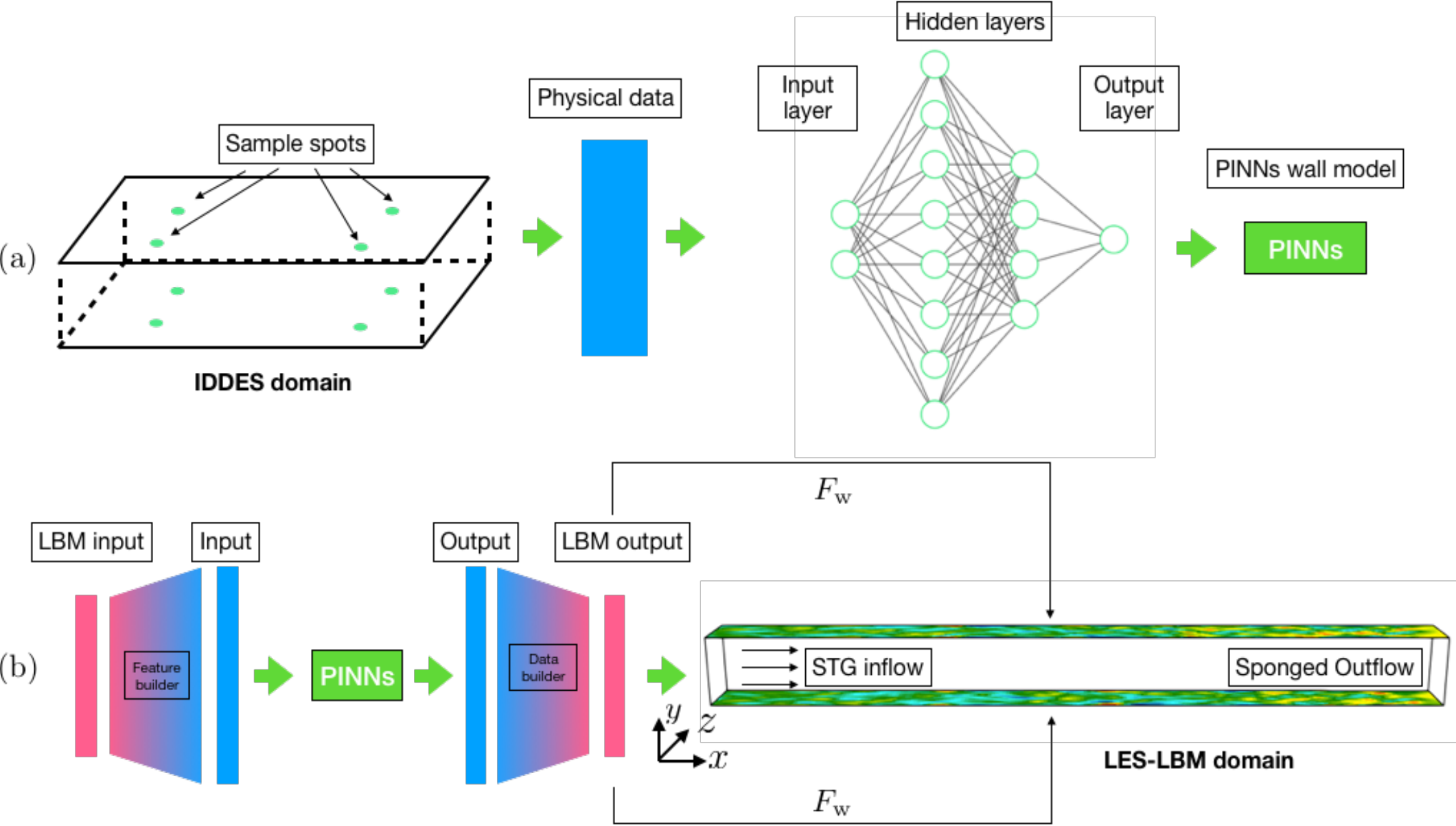}
\caption{Diagram of the physics-informed neural network architecture and implementation for the lattice Boltzmann method trained by IDDES data at $Re_{\tau}=5200$. Panel (a): the training data is obtained from the upper and lower near-wall region of a periodic channel flow simulation. We use NN to predict the shear velocity which can be used by the wall model. Panel (b): LES based LBM channel flow simulation is performed, on the first layer of the wall, we use the LBM data as the input. With the help of the pre-trained PINNs wall model, we can predict LBM shear velocity to calculate the resistance wall force $F_{\text{w}}$ for the wall model.}\label{fig:diagram} 
\end{figure}
As shown in \cref{fig:diagram} (a), we uniformly sample the upper and lower panels of the IDDES periodic channel flow~\cite{shur2008hybrid} at grid points where $\mathrm{y}^+ < 200$. Here, $\mathrm{y}^+$ represents the dimensionless wall-normal distance, defined as $\mathrm{y}^+ = \frac{y u_\tau}{\nu}$, where $\nu$ denotes the kinematic viscosity and $\mathrm{u}_{\tau}$ is the shear velocity. We ensure consistency in the sampling locations across different instantaneous snapshots for the channel flow simulations. The IDDES simulation of turbulent channel flow is conducted at $Re_{\tau} = 5200$. Owing to the varying grid density near the wall, the sampled data may range from dense to sparse. Our primary objective is the accurate prediction of the shear velocity, $\mathrm{u}_{\tau}$, to reliably estimate wall velocities. It is crucial to recognize that the grids used in IDDES are distinct from those in lattice Boltzmann configurations. In contrast to conventional CFD methodologies, the LBM utilizes a uniform grid structure and is formulated with dimensionless variables. This distinction in grid structure necessitates precise capture of the physics inherent in the near-wall region. Figure \ref{fig:diagram} (b) illustrates the pipeline for applying a neural network wall model to LBM turbulent channel flow. The coordinates of the LBM channel flow are denoted as $x, y, z$, corresponding to the stream-wise, vertical, and span-wise directions, respectively. Turbulent flow is initiated with the synthetic turbulence generation (STG) method at the inlet, while the outlet features a sponge region~\cite{xue2022synthetic}. The domain dimensions of the turbulent channel flow are $L_x \times L_y \times L_z$, measuring $320 \times 20 \times 16$ lattice Boltzmann units (LBU). The upper and lower boundaries of the channel employ a free-slip condition. The near-wall region is modelled using a volume force predicted by neural networks. The process involves converting LBM data into PINNs feature input with the help of a feature builder, followed by predicting the feature output of shear velocity. Subsequently, this prediction is converted back into LBM data with the help of a data builder to compute the resistive volume force, thereby correcting near-wall velocities. In the feature builder, a normalizer and LBM-to-physical unit converter are utilized to ensure that input values consistently fall within the range $\left [ 0, 1 \right ]$. The data builder consists of a de-normalizer and a physical-to-LBM unit converter. A detailed description of the NN model is provided in~\cref{sec:method-pinns} and \cref{sec:method-training}. 
\begin{figure}
\centering
\includegraphics[width=1\textwidth]{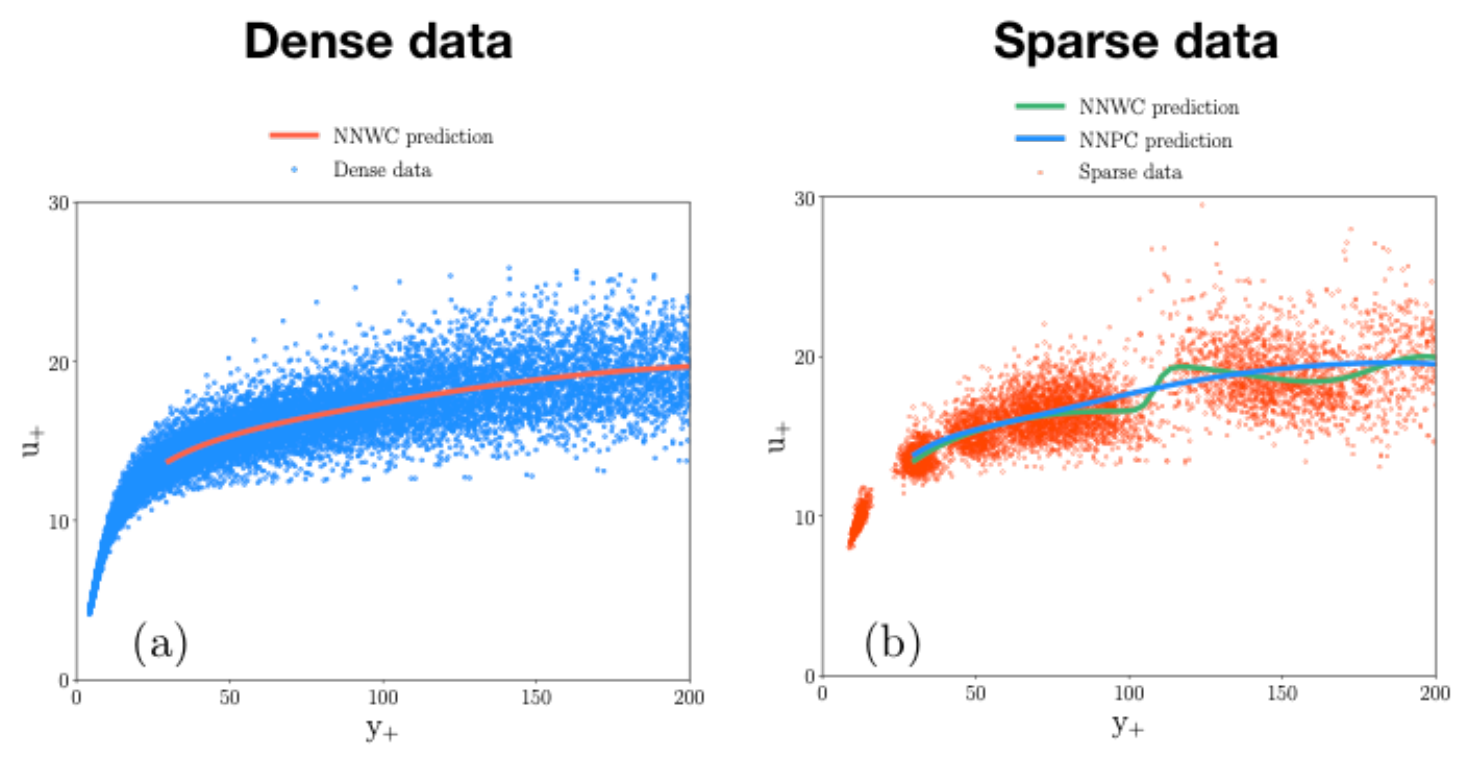}
\caption{Diagram of NN learning the law of wall at $\mathrm{y}^+<200$ with IDDES data. Panel (a): demonstrates dense data configuration (blue dots) with $\mathrm{u}^+$ as function of $\mathrm{y}^+$. NNWC prediction is illustrated by the orange line. Panel (b): demonstrates sparse data configuration (orange dots) comparing NNWC and NNPC predictions.}
\label{fig:densevssparse} 
\end{figure}

\subsection{Dense-data wall model vs sparse-data wall model}

Due to varying grid density near the wall in IDDES data, ``gaps'' may exist between data points. As shown in~\cref{fig:densevssparse} (a), blue dots represent dense data, indicating densely arranged grids below $\mathrm{y}^+<200$ near the wall. The application of neural networks (NN) to learn wall functions from data is demonstrated. The orange line confirms that NNs accurately capture the wall functions at $\mathrm{y}^+ < 200$. This NN model, which requires no data correction, is thus referred to as the ``NNWC'' model (Neural Networks Without PDF Correction). The NNWC model's predictions correspond closely with the IDDES data. In~\cref{fig:densevssparse} (b), orange dots represent sparsely configured grids near the wall, enhancing computational efficiency. However, the direct application of the NNWC model to sparse data, as depicted by the dark green line, risks overfitting and creating a staggered wall function. To mitigate this, we apply PDF corrections, minimizing the impact of ``gap'' regions between data groups during training (see~\cref{sec:method-training}). As shown by the blue line, the NNPC model (Neural Networks with PDF Correction) can precisely capture the law of the wall even with sparse training data.\\ 

\begin{figure}[t]
\centering
\includegraphics[width=1\textwidth]{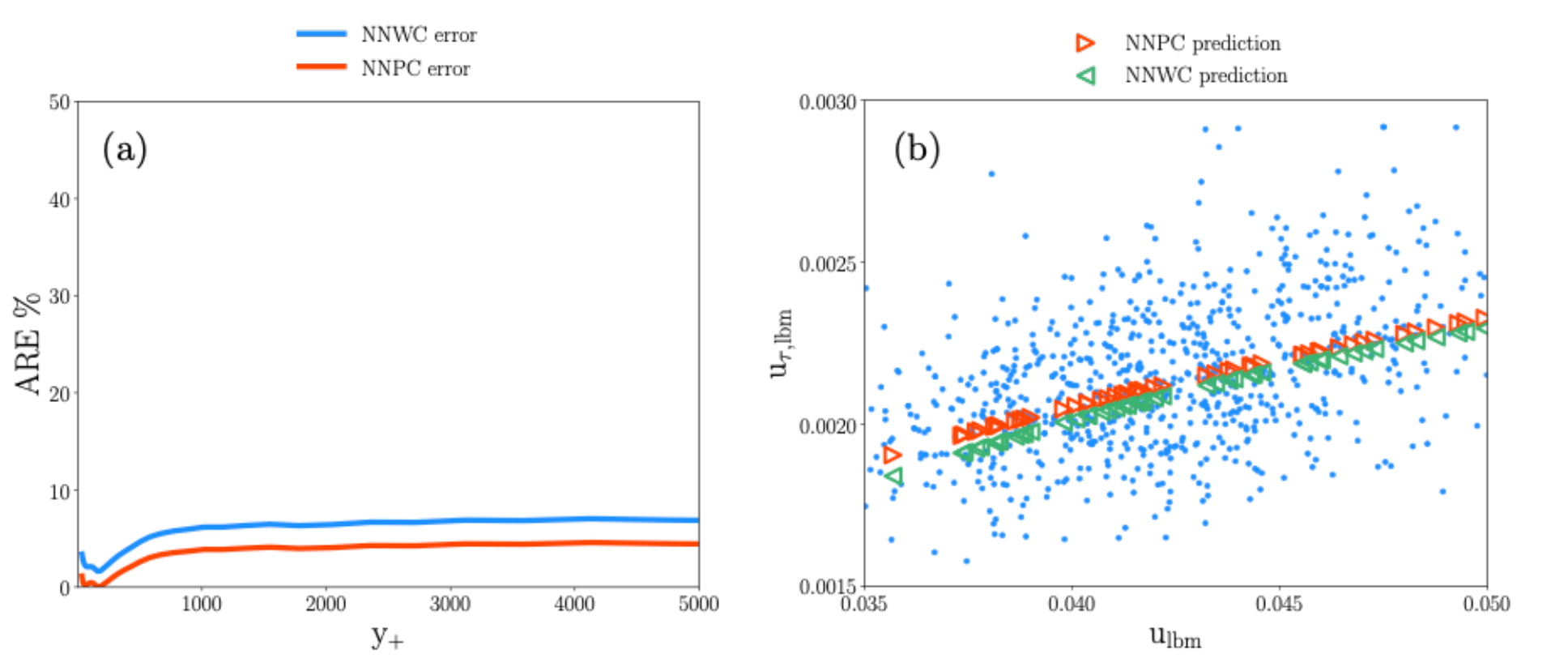}
\caption{Panel (a): absolute relative error (ARE) for $\mathrm{u}_{\tau}$ prediction based on IDDES simulations for $\mathrm{y}^+$ up to $5200$ with or without PDF correction. Panel (b): LBM data prediction on shear velocity $\mathrm{u}_{\tau, \text{lbm}}$  with input of LBM velocity $\mathrm{u}_{\text{lbm}}$ at $\mathrm{y}^+ = 260$ for the channel flow simulation at $Re_{\tau} = 5200$. Blue dots represents the LBM data, the NNPC and NNWC model predictions are represented by the red and green triangular dots respectively.} 
\label{fig:ARE}
\end{figure}

Figure \ref{fig:ARE} (a) depicts the Absolute Relative Error (ARE) in predicting shear velocity, $\mathrm{u}_\tau$, using neural networks on IDDES data, comparing both the NNWC and NNPC models. The blue curve illustrates the ARE of the NNWC model's predictions for dense data. This error notably decreases to $2\%$ around $\mathrm{y}^+ \approx 200$, reflecting the model's training on datasets below $\mathrm{y}^+ = 200$. Beyond this point, the ARE rises to $6\%$ but remains stable up to $\mathrm{y}^+ = 5000$. Conversely, the orange curve represents the ARE of the NNPC model for sparse data. With the application of PDF correction, the ARE drops below $1\%$ near $\mathrm{y}^+ \approx 200$. Despite a slight increase in ARE beyond $\mathrm{y}^+ = 200$, it stabilizes at around $4\%$ for $\mathrm{y}^+ \leq 5000$. The NNPC model demonstrates an accuracy improvement of $2\%$ over the NNWC model. In \cref{fig:ARE} (b), orange triangles represent NNPC predictions, while green triangles are indicative of NNWC predictions. Both models exhibit good alignment with the LBM data. Based on the findings in \cref{fig:densevssparse} and \cref{fig:ARE}, the NNWC model is selected for interpolation and extrapolation tests in LBM channel flow simulations.

\subsection{Interpolation validation for the neural network wall model up to $Re_{\tau} = 5200$}
\begin{figure}
\centering
\includegraphics[width=1\textwidth]{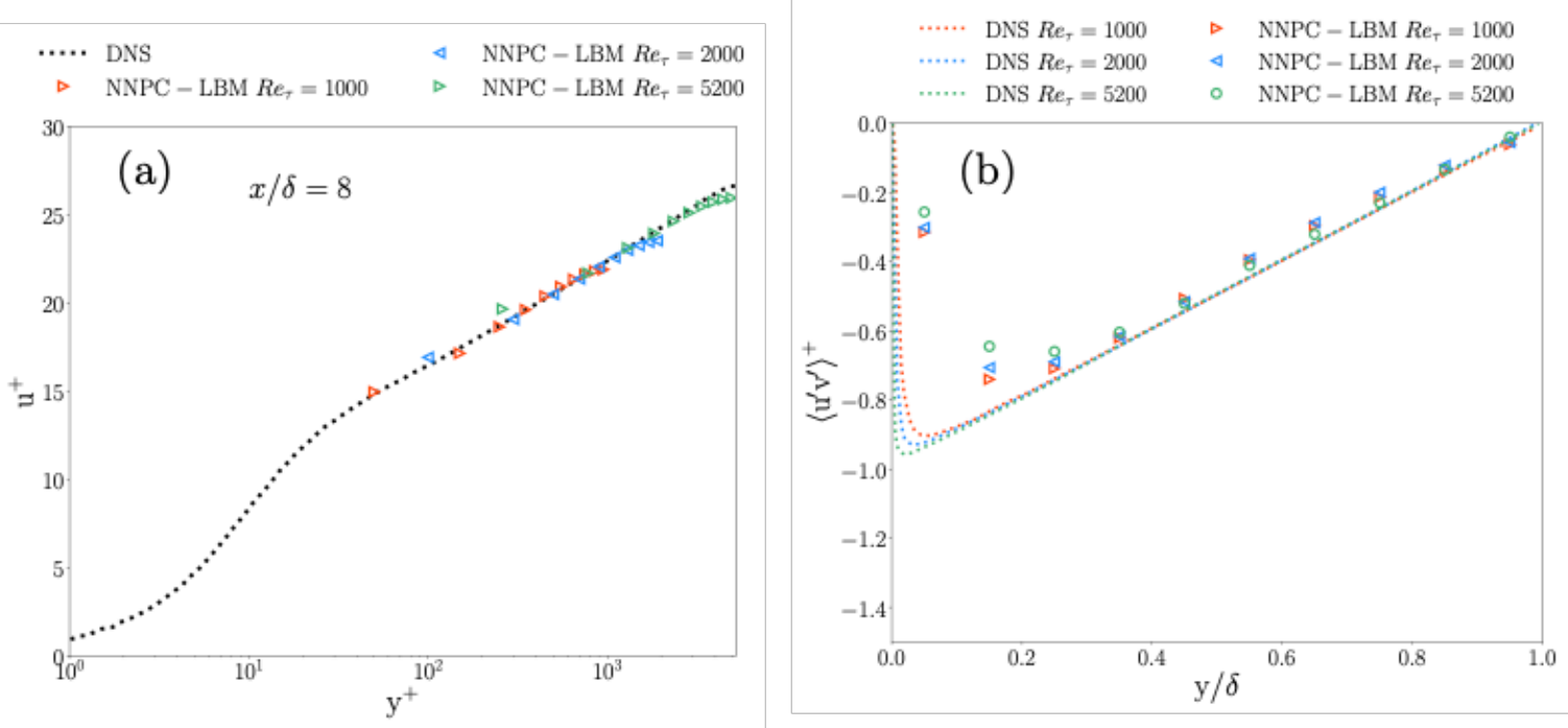}
\caption{Interpolation validation for turbulent channel flow up to $Re_{\tau} = 5200$. Panel (a): Mean velocity, $\mathrm{u}^{+}$, as function of $\mathrm{y}^{+}$ at $Re_{\tau} = 1000, 2000, 5200$. Panel (b): $\left \langle \mathrm{u}^\prime\mathrm{v}^\prime \right \rangle^+$ as function of $\mathrm{y/\delta}$ at $Re_{\tau} = 1000, 2000, 5200$} 
\label{fig:interpolation}
\end{figure}

The turbulent channel flow serves as the validation case for the neural network wall model. In this setup, the channel flow employs a LBM-based synthetic turbulence generator at the inlet, as described in~\cite{xue2022synthetic}, and incorporates a sponge layer at the outlet. The volume force, $F_{\text{w}}(\mathbf{x}, t)$, exerted on the first layer of cells near the wall, can be characterized at location $\mathbf{x}$ as

\begin{equation}
\label{eq:force_wm}
F_{\text{w}}(\mathbf{x}, t) = - \tau_{\text{w}}(\mathbf{x}, t) A,
\end{equation}
where $\tau_{\text{w}}$ can be obtained by shear velocity and density: $\tau_{\text{w}}(\mathbf{x}, t) = u_{\tau, \text{lbm}}^2(\mathbf{x}, t) \rho(\mathbf{x}, t)$. $A$ is the acting area. The height of the channel is set to 20 LBU. Consequently, the $\mathrm{y}^+$ values of the first cell layer for friction Reynolds numbers $Re_{\tau} = 1000$, $2000$, and $5200$ are approximately 50, 100, and 260, respectively. These results are benchmarked against DNS data from~\cite{hoyas2006scaling, lee2015direct}. Owing to the rapid convergence characteristics of the turbulence generator~\cite{xue2022synthetic}, simulations are executed at various friction Reynolds numbers. Each simulation runs for a total duration of 12 domain-through-times ($12T$), with statistical analyses commencing post $2T$. The cross-section sampling is performed at $x/\delta = 8$ to ensure robust statistical properties of the turbulence, where $\delta$ is the turbulent boundary layer thickness and is equal to $\delta = 10$ LBU for the channel flow case.
 Figure \ref{fig:interpolation} (a) illustrates the mean flow velocity $\mathrm{u}^+$ of LBM turbulent flow as a function of $\mathrm{y}^+$. The red, blue, and green dots correspond to LBM simulations at $Re_{\tau} = 1000, 2000, 5200$, respectively. The black dotted line represents the DNS data at $Re_{\tau} = 5200$. The results of our LBM NNPC-based wall model demonstrate good alignment with the DNS data. Subsequently, the turbulent shear stress $\left \langle \mathrm{u}^\prime\mathrm{v}^\prime \right \rangle^+$ is compared with the DNS reference data. Figure \ref{fig:interpolation} (b) displays $\left \langle \mathrm{u}^\prime\mathrm{v}^\prime \right \rangle^+$ as a function of $\mathrm{y/\delta}$ for $Re_{\tau} = 1000, 2000, 5200$. The red, blue, and green dotted lines represent the DNS references~\cite{hoyas2006scaling, lee2015direct}, while the corresponding coloured points denote the LBM NNPC-based wall model results. Discrepancies are observed in the first three cells near the wall, attributable to coarse-grained resolution issues, given that the channel height comprises only 20 lattice units. Beyond the fourth grid cell, the LBM results closely match the DNS data, indicating the efficacy of our wall model in accurately capturing the physics of the turbulent stress tensor in channel flow.

\subsection{Extrapolation studies for the neural network wall model up to $Re_{\tau} = 1.0 \times 10^6$}
To demonstrate the generality of our NNPC-based wall model, we extended its application to friction Reynolds numbers $Re_{\tau} = 1.0 \times 10^5$ and $Re_{\tau} = 1.0 \times 10^6$, which are two orders of magnitude higher than those considered in previous LBM-base wall model studies~\cite{malaspinas2014wall, haussmann2019large, maeyama2020unsteady, pasquali2020near, wilhelm2021new}. For channel flow simulations at these elevated Reynolds numbers, the first cell layer of the NNPC-based model corresponds to approximately $\mathrm{y}^+= 5000$ and $\mathrm{y}^+= 50000$, respectively, significantly exceeding the range of the initial training dataset.

\begin{figure}
\centering
\includegraphics[width=1\textwidth]{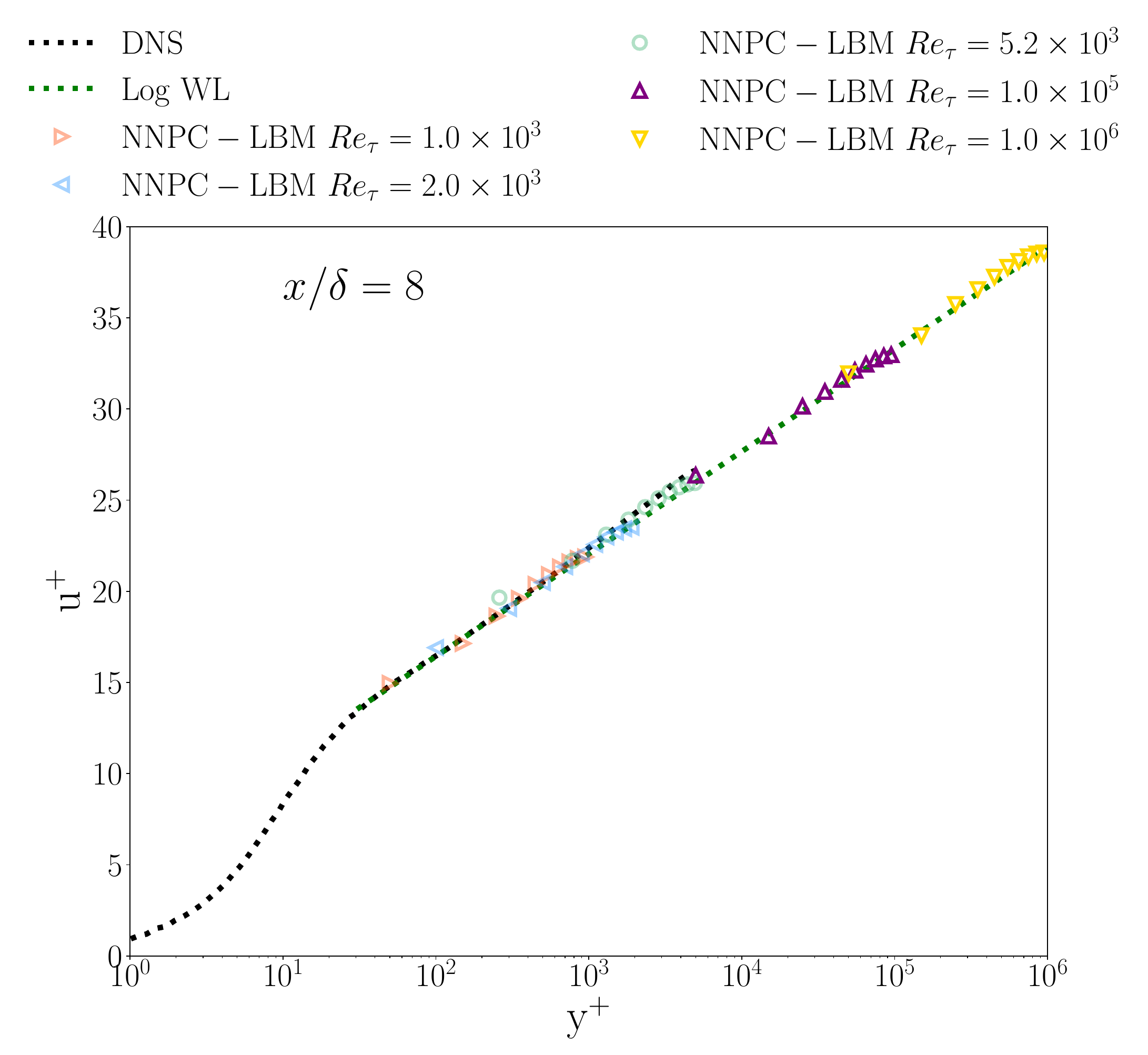}
\caption{Extrapolation validation: mean velocity $\mathrm{u}^{+}$ as function of $\mathrm{y}^{+}$ at $Re_{\tau} = 10^5$ (purple dots) and $Re_{\tau} = 10^6$ (yellow dots). The green dotted line is the log wall law satisfying~\cref{eq:logwl}, for which DNS data is only available up to $Re_{\tau} = 5200$.} 
\label{fig:extrapolation}
\end{figure}

Figure~\ref{fig:extrapolation} presents $\mathrm{u}^+$ as a function of $\mathrm{y}^+$ for $Re_{\tau} = 1.0 \times 10^5$ and $1.0 \times 10^6$. The black dotted line represents the DNS reference at $Re_{\tau} = 5200$. Due to the absence of DNS data for these higher Reynolds numbers, the green dotted line depicts the universal logarithmic wall law (Log WL), which is expressed as

\begin{equation}
\label{eq:logwl}
\mathrm{u}^+ = 1/\kappa \mathrm{log}\mathrm{y}^+ + B,
\end{equation}
where $\kappa$ is chosen to be $0.42$ and $B$ is set to $5.2$. The purple and yellow dots represent the LBM-based NNPC wall model results at $Re_{\tau} = 1.0 \times 10^5$ and $Re_{\tau} = 1.0 \times 10^6$, respectively. Results from the channel flow simulations under $Re_{\tau} = 5200$ are also included in the figure, depicted in faded colours for comparative purposes. Although our neural network model is primarily guided by physical data, it aligns remarkably well with the logarithmic wall law up to $Re_{\tau} = 1.0 \times 10^6$. This concordance highlights the versatility and effectiveness of our NNPC-based wall model within the LBM framework.

\section{Concluding remarks}\label{sec:conclusions}
This paper introduces a physics-informed neural network wall model tailored for the lattice Boltzmann method. The model was trained using IDDES channel flow simulations at $Re_{\tau} = 5200$. Due to grid sparsity near the wall, the training dataset encompasses both dense and sparse data configurations below $\mathrm{y}^+=200$. Implementing PDF corrections leads to a $2\%$ reduction in error compared to models without such corrections, yielding an approximate absolute relative error of $4\%$ for $\mathrm{y}^+>1000$. We evaluated our NNPC-based wall model in LBM channel flow simulations using STG as the inlet. Despite the coarse-grained nature of LBM channel flow, our simulation results align well with DNS data up to $Re_{\tau} = 5200$. It is noteworthy that previous LBM channel flow studies can only validate the wall model till $Re_{\tau} = 2.0 \times 10^4$~\cite{malaspinas2014wall}. We further validate our NNPC-based LBM wall model with the logarithmic wall law at scale of $Re_{\tau} = 10^6$ where DNS data is not available. This evidence supports the conclusion that our NNPC-based wall model is effective for arbitrarily high friction Reynolds numbers. It should be noted that investigations beyond this range were not conducted as they exceed typical application requirements. Nonetheless, we are confident that the model will adhere to the logarithmic wall law even at higher friction Reynolds numbers. As a concluding remark, our novel NNPC-based wall-model demonstrates effective performance on coarse-grained grids near the wall. It is versatile enough to be applied to flows at arbitrarily high Reynolds numbers, opening up a broad array of aerodynamics-related industrial applications.

\section{Methods}
\subsection{The multiple-relaxation time lattice Boltzmann method}\label{sec:method-lbm}
This study employs a three-dimensional (3D) Lattice Boltzmann model featuring 19 discretized directions, known as the D3Q19 model. The lattice cell is specified by its position $\mathbf{x}$ and time $t$, and is characterized by a discretized velocity set $\mathbf{c}_i$ where $i \in \{0, 1, \ldots, Q-1\}$ with $Q=19$. Macro-scale quantities such as density, momentum, and momentum flux tensors are derived from the distribution function $f_i(\mathbf{x}, t)$, the discrete velocities $\mathbf{c}_{i}$, and the volume acceleration,
 $\mathbf{g}$,:
\begin{equation}\label{eq:density}
\rho(\mathbf{x}, t) = \sum_{i=0}^{Q-1} f_i(\mathbf{x}, t), \\
\end{equation} 
\begin{equation}
\label{eq:momentum}
\rho(\mathbf{x}, t)\mathbf{u}(\mathbf{x}, t) = \sum_{i=0}^{Q-1} f_i(\mathbf{x}, t)\mathbf{c}_{i} + \frac{1}{2}\mathbf{g}\Delta t,
\end{equation}
\begin{equation}
\label{eq:tensor}
\mathbf{\Pi}(\mathbf{x}, t) = \sum_{i=0}^{Q-1} f_i(\mathbf{x}, t)\mathbf{c}_{i}\mathbf{c}_{i}.
\end{equation}
The evolution equation for the distribution functions, accounting for collision and forcing, can be expressed as:
\begin{equation}
\label{eq:lbe}
\mathbf{f}(\mathbf{x}+\mathbf{c}_{i}\Delta t,t+\Delta t) =\mathbf{f}(\mathbf{x}, t) - \mathbf{\Omega} \left[\mathbf{f}(\mathbf{x},t )-\mathbf{f}^{\mbox{ eq}}(\mathbf{x},t )\right] + \mathbf{F}(\mathbf{x}, t) \Delta t,
\end{equation}
where $\mathbf{\Omega}$ denotes the multiple relaxation time (MRT) collision kernel, defined as $\mathbf{\Omega} = \mathbf{M}^{-1}\mathbf{S}\mathbf{M}$\cite{d2002multiple}. The matrix $\mathbf{S}$ is a diagonal matrix comprising relaxation frequencies for different moments, expressed as $\mathbf{S}=diag\{\omega_0, \omega_1, \ldots , \omega_{Q-1}\}$. The matrix $\mathbf{M}$ represents the transformation matrix that converts population space to moment space, derived using the Gram-Schmidt orthogonalization process. In~\cref{eq:lbe}, $\Delta t$ symbolizes the lattice Boltzmann time step, which is standardized to unity. The frequency $\omega_i$ corresponds to the inverse of relaxation time $\tau_i$. It is important to note that we equate $\tau_{k} = \tau_{9}=\tau_{11}=\tau_{13}=\tau_{14}=\tau_{15}$, all of which are associated with the kinematic viscosity $\nu$, which is

\begin{equation}
\label{eq:nu}
\nu = c_s^2\left(\tau_{k}-\frac{1}{2}\right)\Delta t,
\end{equation}
with $c_s$ representing the speed of sound, and $c_s^2$ equating to $\frac{1}{3}$ in Lattice Boltzmann Units (LBU). The other relaxation parameters, which are not governed by the viscosity, are set as follows:

\begin{align}
\label{eq:freeparams}
\omega_{0} &= \omega_{3}=\omega_{5}=\omega_{7}=\omega_{1}=1.0\\
\omega_1 &= 1.19,\\ 
\omega_2 &= \omega_{10} = \omega_{12} = 1.6\\
\omega_{4} &= \omega_{6} = \omega_{8}=1.2\\
\omega_{16} &= \omega_{17} = \omega_{18}=1.98.
\end{align} 
$\mathbf{F}(\mathbf{x}, t)$ in~\cref{eq:lbe} is the vector of $F_i(\mathbf{x}, t)$ which is the force acting on the fluid cell \cite{guo2002discrete}:
\begin{equation}
\label{eq: bodyforce}
F_i(\mathbf{x}, t) = (1 - \frac{\omega_i}{2}) w_i \left [ \frac{\textbf{c}_i - \textbf{u}\left(\mathbf{x},t\right)}{c_s^2} + \frac{\textbf{c}_i \cdot \textbf{u}\left(\mathbf{x},t\right)}{c_s^4}\textbf{c}_i \right] \cdot \mathbf{g}.
\end{equation}

\subsection{Smagorinsky subgrid-scale modelling}\label{sec:method-sgs}
In this part, we summarize the lattice-Boltzmann-based Smagorinsky Subgrid Scale (SGS) LES techniques. Within the LBM framework, the effective viscosity $\nu _{\mathrm{eff}}$ \cite{smagorinsky1963general, hou1994lattice, koda2015lattice} is modelled as the sum of the molecular viscosity, $\nu_0$, and the turbulent viscosity, $\nu_t$:

\begin{equation}
\nu _{\mathrm{eff}}=\nu _0+\nu_t, \hspace{.6in} \nu_t = C_{\mathrm{smag}}\Delta ^2\left |\bar{\mathbf{S}}\right |,
\label{smogrinsky_model}
\end{equation}
where $\left |\mathbf{\bar{S}} \right |$ is the filtered strain rate tensor, $C_{\mathrm{smag}}$ is the Smagorinsky constant, $\Delta$ represents the filter size.

\subsection{Synthetic turbulence generator formulation}\label{sec:method-stg}
The synthetic turbulence generator (STG) requires a velocity field given by a $k-\varepsilon$ Reynolds-Averaged Navier-Stokes (RANS) simulation \cite{abe1994new}. The total velocity $\mathbf{u}_{\text{in}}(\mathbf{x}, t)$ at the inlet is given by

\begin{equation}
\label{eq:u_couple}
\mathbf{u}_{\text{in}}(\mathbf{x}, t) = \mathbf{u}_{\text{RANS}}(\mathbf{x}) + \mathbf{u}'(\mathbf{x}, t),
\end{equation}
where $\mathbf{u}_{\text{RANS}}$ is the velocity vector obtained from a RANS simulation, then the interpolated velocity will be applied on LBM grid in case of grid resolution differences~\cite{xue2022synthetic}. The STG generates the velocity fluctuations $\mathbf{u}'(\mathbf{x}, t)$ at the cell $\mathbf{x}$ at time $t$:
\begin{equation}
\label{eq:v_fluc}
\mathbf{u}'(\mathbf{x}, t)=a_{\alpha\beta}\mathbf{v}'(\mathbf{x}, t).
\end{equation}
The time-averaged velocity fluctuation is zero, i.e., $\langle \mathbf{u}'(\mathbf{x}, t)\rangle = 0$. The term $a_{\alpha\beta}$ represents the Cholesky decomposition of the Reynolds stress tensor. The fluctuations $\mathbf{v}'(\mathbf{x}, t)$ are imposed by $N$ Fourier modes. Detailed descriptions of this process are available in Xue and co-authors~\cite{xue2022synthetic}.

\subsection{IDDES database preparation}\label{sec:method-iddes}
The finite volume code pyCALC-LES \cite{davidson2021pycalc} is utilised, which is implemented in Python and fully vectorised. Spatial discretisation for the momentum equations employs a second-order central difference scheme, while temporal discretisation utilizes the Crank-Nicolson scheme. The evolution of the $k-\varepsilon$ RANS turbulence model adopts a hybrid central/upwind scheme, complemented by a first-order fully implicit time discretisation approach. The discretised equations, forming a sparse-matrix system, are solved on the GPU using the Algebraic Multi-Grid library pyAMGx \cite{olson2018pyamg}. For turbulent data generation, the dimensions of the channel flow in the streamwise, wall-normal, and spanwise directions are set to $3.2 \times 2 \times 1.6$ meters, respectively. The computational mesh comprises $96 \times 96 \times 96$ cells, and the Reynolds number based on the friction velocity, $Re_{\tau}$, is set to 5200. 

\subsection{Physics-informed neural networks for wall model}\label{sec:method-pinns}
We use a PINN model that contains several fully connected hidden layers with the $tanh$ function as activation function. Generally, the input is reformulated so that it provides crucial physics and helps to find a convergence in the optimization process when training the model.

The detailed description of our NN models are listed in Tab.\ref{Tab:NN architechture}. The NNPC and NNWC models were both tested on predicting $\mathrm{y}^+$ and $\mathrm{u}^+$. However, to ensure we identify a model with the generality to predict shear velocity $\mathrm{u}_{\tau}$, we need actual physics-informed input. We choose $\mathrm{u}/(1000\mathrm{y})$ and $\mathrm{y}/\mathrm{y}_{\text{ref}}/\mathrm{u}$ to learn the log scale slope of the log wall law in~\cref{eq:logwl}. The model outputs a continuous value in the interval [0,1], which is linearly mapped to [ $\mathrm{u}_{\tau, \text{min}}$, $\mathrm{u}_{\tau, \text{max}}$] for further application.

\begin{table}
    \centering
         \begin{tabular}{|>{\centering\arraybackslash}p{2cm}|>{\centering\arraybackslash}p{3cm}|>{\centering\arraybackslash}p{4cm}|>{\centering\arraybackslash}p{3cm}|}
        \hline \hline 
        NN model & HL size & Input & Output \\
        \hline 
        NNPC \& NNWC & $(9,9)$ & $\mathrm{y}^+$ & $\mathrm{u}^+$ \\
        \hline 
        NNPC & $(9,8,5)$ & $(\frac{\mathrm{u}}{1000\mathrm{y}},\frac{\mathrm{y}/\mathrm{y}_{\text{ref}}}{\mathrm{u}})$ & $\mathrm{u}_{\tau}$ \\
        \hline \hline
    \end{tabular}
    \caption{Details of the neural networks for law of wall prediction and shear velocity $\mathrm{u}_{\tau}$ prediction. Here NN denotes neural network and HL denotes hidden layer. The tabulated hidden layer size contains the number of neurons within each hidden layer.}
    \label{Tab:NN architechture}
\end{table}

\subsection{Neural network training}\label{sec:method-training}
We trained the PINNs model in a supervised fashion, with a loss function $\mathcal{L}$, consisting of $L_2$ norm and a weight function. The model is trained by a sample of size $N$, with a predicted vector $\mathbf{X}_{\mathrm{PINNs}}$ and target vector $\mathbf{Y}$, the total loss yields: 
\begin{equation}
    \mathcal{L}= \frac{1}{N} \sum_{n=1}^N w_{\mathrm{loss}}\left \|\mathbf{Y}-\mathbf{X}_{\mathrm{PINNs}}\right \|^2.
\end{equation}
Here the weight function $w_{\mathrm{loss}}=1/\left[P_r\left(\mathbf{X}_{\mathrm{PINNs}}\right)+\epsilon\right]$ compensates for the gaps in data points (as shown in Fig.~\cref{fig:densevssparse} (a)). The constant $\epsilon =10^{-6}$ avoids infinite value of the weight function. To provide a general solution to deal with the abitrary ``gap" on data sample, the weight function is also learned during the training process. 
An $M$-dimensional input data with $N$ sample points is firstly normalized and linearly projected into interval [0,1]. Here, the $M$ is number of inputs which is set to $2$. The data range across each dimension is discretized into $J$ uniform segments. Therefore, the likelihood of a data point being located in the $j$th segment of the $i$th dimension is quantified as

\begin{equation}
    P(i,j_\mathrm{i})=\frac{N_{\mathrm{i,j}}}{N},
\end{equation}
where $N_{i,j}$ denotes the number of training data falls in the $j$th segment in $i$th dimension. We use the expression
\begin{equation}
    P_r(j_1,j_2...j_\mathrm{M})=\prod_{i=1}^{M} P(i,j_\mathrm{i})
\end{equation}
to denote the relative possibility of a data point that falls in the $(j_1,j_2...j_\mathrm{M})$ segment. We pragmatically used $J = 100$ in the training phase.

The sparse dataset is uniformly sampled from a total of 10,000 input-output pairs. This dataset comprises 2,000 points that are uncorrelated in both time and space, gathered from five layers of grids from the IDDES channel flow simulations located under $\mathrm{y}^+<200$. For the model trained on dense data, the dataset consists of 19,600 input-output pairs, sourced from the grids under $\mathrm{y}^+<200$. We adopted a training-to-test ratio of 80\% to 20\%.

\bibliography{prex}

\hyphenation{Post-Script Sprin-ger}
\begin{thebibliography}{56}%
\makeatletter
\providecommand \@ifxundefined [1]{%
 \@ifx{#1\undefined}
}%
\providecommand \@ifnum [1]{%
 \ifnum #1\expandafter \@firstoftwo
 \else \expandafter \@secondoftwo
 \fi
}%
\providecommand \@ifx [1]{%
 \ifx #1\expandafter \@firstoftwo
 \else \expandafter \@secondoftwo
 \fi
}%
\providecommand \natexlab [1]{#1}%
\providecommand \enquote  [1]{``#1''}%
\providecommand \bibnamefont  [1]{#1}%
\providecommand \bibfnamefont [1]{#1}%
\providecommand \citenamefont [1]{#1}%
\providecommand \href@noop [0]{\@secondoftwo}%
\providecommand \href [0]{\begingroup \@sanitize@url \@href}%
\providecommand \@href[1]{\@@startlink{#1}\@@href}%
\providecommand \@@href[1]{\endgroup#1\@@endlink}%
\providecommand \@sanitize@url [0]{\catcode `\\12\catcode `\$12\catcode
  `\&12\catcode `\#12\catcode `\^12\catcode `\_12\catcode `\%12\relax}%
\providecommand \@@startlink[1]{}%
\providecommand \@@endlink[0]{}%
\providecommand \url  [0]{\begingroup\@sanitize@url \@url }%
\providecommand \@url [1]{\endgroup\@href {#1}{\urlprefix }}%
\providecommand \urlprefix  [0]{URL }%
\providecommand \Eprint [0]{\href }%
\providecommand \doibase [0]{http://dx.doi.org/}%
\providecommand \selectlanguage [0]{\@gobble}%
\providecommand \bibinfo  [0]{\@secondoftwo}%
\providecommand \bibfield  [0]{\@secondoftwo}%
\providecommand \translation [1]{[#1]}%
\providecommand \BibitemOpen [0]{}%
\providecommand \bibitemStop [0]{}%
\providecommand \bibitemNoStop [0]{.\EOS\space}%
\providecommand \EOS [0]{\spacefactor3000\relax}%
\providecommand \BibitemShut  [1]{\csname bibitem#1\endcsname}%
\let\auto@bib@innerbib\@empty
\bibitem [{\citenamefont {Port{\'e}-Agel}\ \emph {et~al.}(2011)\citenamefont
  {Port{\'e}-Agel}, \citenamefont {Wu}, \citenamefont {Lu},\ and\ \citenamefont
  {Conzemius}}]{porte2011large}%
  \BibitemOpen
  \bibfield  {author} {\bibinfo {author} {\bibfnamefont {F.}~\bibnamefont
  {Port{\'e}-Agel}}, \bibinfo {author} {\bibfnamefont {Y.-T.}\ \bibnamefont
  {Wu}}, \bibinfo {author} {\bibfnamefont {H.}~\bibnamefont {Lu}}, \ and\
  \bibinfo {author} {\bibfnamefont {R.~J.}\ \bibnamefont {Conzemius}},\
  }\bibfield  {title} {\enquote {\bibinfo {title} {Large-eddy simulation of
  atmospheric boundary layer flow through wind turbines and wind farms},}\
  }\href@noop {} {\bibfield  {journal} {\bibinfo  {journal} {Journal of Wind
  Engineering and Industrial Aerodynamics}\ }\textbf {\bibinfo {volume} {99}},\
  \bibinfo {pages} {154--168} (\bibinfo {year} {2011})}\BibitemShut {NoStop}%
\bibitem [{\citenamefont {Mehta}\ \emph {et~al.}(2014)\citenamefont {Mehta},
  \citenamefont {Van~Zuijlen}, \citenamefont {Koren}, \citenamefont
  {Holierhoek},\ and\ \citenamefont {Bijl}}]{mehta2014large}%
  \BibitemOpen
  \bibfield  {author} {\bibinfo {author} {\bibfnamefont {D.}~\bibnamefont
  {Mehta}}, \bibinfo {author} {\bibfnamefont {A.}~\bibnamefont {Van~Zuijlen}},
  \bibinfo {author} {\bibfnamefont {B.}~\bibnamefont {Koren}}, \bibinfo
  {author} {\bibfnamefont {J.}~\bibnamefont {Holierhoek}}, \ and\ \bibinfo
  {author} {\bibfnamefont {H.}~\bibnamefont {Bijl}},\ }\bibfield  {title}
  {\enquote {\bibinfo {title} {{L}arge {E}ddy {S}imulation of wind farm
  aerodynamics: A review},}\ }\href@noop {} {\bibfield  {journal} {\bibinfo
  {journal} {Journal of Wind Engineering and Industrial Aerodynamics}\ }\textbf
  {\bibinfo {volume} {133}},\ \bibinfo {pages} {1--17} (\bibinfo {year}
  {2014})}\BibitemShut {NoStop}%
\bibitem [{\citenamefont {Chapman}(1979)}]{chapman1979computational}%
  \BibitemOpen
  \bibfield  {author} {\bibinfo {author} {\bibfnamefont {D.~R.}\ \bibnamefont
  {Chapman}},\ }\bibfield  {title} {\enquote {\bibinfo {title} {Computational
  aerodynamics development and outlook},}\ }\href@noop {} {\bibfield  {journal}
  {\bibinfo  {journal} {AIAA journal}\ }\textbf {\bibinfo {volume} {17}},\
  \bibinfo {pages} {1293--1313} (\bibinfo {year} {1979})}\BibitemShut {NoStop}%
\bibitem [{\citenamefont {Choi}\ and\ \citenamefont
  {Moin}(2012)}]{choi2012grid}%
  \BibitemOpen
  \bibfield  {author} {\bibinfo {author} {\bibfnamefont {H.}~\bibnamefont
  {Choi}}\ and\ \bibinfo {author} {\bibfnamefont {P.}~\bibnamefont {Moin}},\
  }\bibfield  {title} {\enquote {\bibinfo {title} {Grid-point requirements for
  large eddy simulation: Chapman?s estimates revisited},}\ }\href@noop {}
  {\bibfield  {journal} {\bibinfo  {journal} {Physics of Fluids}\ }\textbf
  {\bibinfo {volume} {24}},\ \bibinfo {pages} {011702} (\bibinfo {year}
  {2012})}\BibitemShut {NoStop}%
\bibitem [{\citenamefont {Yang}\ and\ \citenamefont
  {Griffin}(2021)}]{yang2021grid}%
  \BibitemOpen
  \bibfield  {author} {\bibinfo {author} {\bibfnamefont {X.~I.}\ \bibnamefont
  {Yang}}\ and\ \bibinfo {author} {\bibfnamefont {K.~P.}\ \bibnamefont
  {Griffin}},\ }\bibfield  {title} {\enquote {\bibinfo {title} {Grid-point and
  time-step requirements for direct numerical simulation and large-eddy
  simulation},}\ }\href@noop {} {\bibfield  {journal} {\bibinfo  {journal}
  {Physics of Fluids}\ }\textbf {\bibinfo {volume} {33}},\ \bibinfo {pages}
  {015108} (\bibinfo {year} {2021})}\BibitemShut {NoStop}%
\bibitem [{\citenamefont {Schumann}(1975)}]{schumann1975subgrid}%
  \BibitemOpen
  \bibfield  {author} {\bibinfo {author} {\bibfnamefont {U.}~\bibnamefont
  {Schumann}},\ }\bibfield  {title} {\enquote {\bibinfo {title} {Subgrid scale
  model for finite difference simulations of turbulent flows in plane channels
  and annuli},}\ }\href@noop {} {\bibfield  {journal} {\bibinfo  {journal}
  {Journal of computational physics}\ }\textbf {\bibinfo {volume} {18}},\
  \bibinfo {pages} {376--404} (\bibinfo {year} {1975})}\BibitemShut {NoStop}%
\bibitem [{\citenamefont {Park}\ and\ \citenamefont
  {Moin}(2014)}]{park2014improved}%
  \BibitemOpen
  \bibfield  {author} {\bibinfo {author} {\bibfnamefont {G.~I.}\ \bibnamefont
  {Park}}\ and\ \bibinfo {author} {\bibfnamefont {P.}~\bibnamefont {Moin}},\
  }\bibfield  {title} {\enquote {\bibinfo {title} {An improved dynamic
  non-equilibrium wall-model for large eddy simulation},}\ }\href@noop {}
  {\bibfield  {journal} {\bibinfo  {journal} {Physics of Fluids}\ }\textbf
  {\bibinfo {volume} {26}} (\bibinfo {year} {2014})}\BibitemShut {NoStop}%
\bibitem [{\citenamefont {Larsson}\ \emph {et~al.}(2016)\citenamefont
  {Larsson}, \citenamefont {Kawai}, \citenamefont {Bodart},\ and\ \citenamefont
  {Bermejo-Moreno}}]{larsson2016large}%
  \BibitemOpen
  \bibfield  {author} {\bibinfo {author} {\bibfnamefont {J.}~\bibnamefont
  {Larsson}}, \bibinfo {author} {\bibfnamefont {S.}~\bibnamefont {Kawai}},
  \bibinfo {author} {\bibfnamefont {J.}~\bibnamefont {Bodart}}, \ and\ \bibinfo
  {author} {\bibfnamefont {I.}~\bibnamefont {Bermejo-Moreno}},\ }\bibfield
  {title} {\enquote {\bibinfo {title} {Large eddy simulation with modeled
  wall-stress: recent progress and future directions},}\ }\href@noop {}
  {\bibfield  {journal} {\bibinfo  {journal} {Mechanical Engineering Reviews}\
  }\textbf {\bibinfo {volume} {3}},\ \bibinfo {pages} {15--00418} (\bibinfo
  {year} {2016})}\BibitemShut {NoStop}%
\bibitem [{\citenamefont {Bose}\ and\ \citenamefont
  {Park}(2018)}]{bose2018wall}%
  \BibitemOpen
  \bibfield  {author} {\bibinfo {author} {\bibfnamefont {S.~T.}\ \bibnamefont
  {Bose}}\ and\ \bibinfo {author} {\bibfnamefont {G.~I.}\ \bibnamefont
  {Park}},\ }\bibfield  {title} {\enquote {\bibinfo {title} {Wall-modeled
  large-eddy simulation for complex turbulent flows},}\ }\href@noop {}
  {\bibfield  {journal} {\bibinfo  {journal} {Annual Review of Fluid
  Mechanics}\ }\textbf {\bibinfo {volume} {50}},\ \bibinfo {pages} {535--561}
  (\bibinfo {year} {2018})}\BibitemShut {NoStop}%
\bibitem [{\citenamefont {Brunton}, \citenamefont {Noack},\ and\ \citenamefont
  {Koumoutsakos}(2020)}]{brunton2020machine}%
  \BibitemOpen
  \bibfield  {author} {\bibinfo {author} {\bibfnamefont {S.~L.}\ \bibnamefont
  {Brunton}}, \bibinfo {author} {\bibfnamefont {B.~R.}\ \bibnamefont {Noack}},
  \ and\ \bibinfo {author} {\bibfnamefont {P.}~\bibnamefont {Koumoutsakos}},\
  }\bibfield  {title} {\enquote {\bibinfo {title} {Machine learning for fluid
  mechanics},}\ }\href@noop {} {\bibfield  {journal} {\bibinfo  {journal}
  {Annual Review of Fluid Mechanics}\ }\textbf {\bibinfo {volume} {52}},\
  \bibinfo {pages} {477--508} (\bibinfo {year} {2020})}\BibitemShut {NoStop}%
\bibitem [{\citenamefont {Yin}\ \emph {et~al.}(2014)\citenamefont {Yin},
  \citenamefont {Ding}, \citenamefont {Xie},\ and\ \citenamefont
  {Luo}}]{yin2014review}%
  \BibitemOpen
  \bibfield  {author} {\bibinfo {author} {\bibfnamefont {S.}~\bibnamefont
  {Yin}}, \bibinfo {author} {\bibfnamefont {S.~X.}\ \bibnamefont {Ding}},
  \bibinfo {author} {\bibfnamefont {X.}~\bibnamefont {Xie}}, \ and\ \bibinfo
  {author} {\bibfnamefont {H.}~\bibnamefont {Luo}},\ }\bibfield  {title}
  {\enquote {\bibinfo {title} {A review on basic data-driven approaches for
  industrial process monitoring},}\ }\href@noop {} {\bibfield  {journal}
  {\bibinfo  {journal} {IEEE Transactions on Industrial electronics}\ }\textbf
  {\bibinfo {volume} {61}},\ \bibinfo {pages} {6418--6428} (\bibinfo {year}
  {2014})}\BibitemShut {NoStop}%
\bibitem [{\citenamefont {Gopalakrishnan}\ \emph {et~al.}(2017)\citenamefont
  {Gopalakrishnan}, \citenamefont {Khaitan}, \citenamefont {Choudhary},\ and\
  \citenamefont {Agrawal}}]{gopalakrishnan2017deep}%
  \BibitemOpen
  \bibfield  {author} {\bibinfo {author} {\bibfnamefont {K.}~\bibnamefont
  {Gopalakrishnan}}, \bibinfo {author} {\bibfnamefont {S.~K.}\ \bibnamefont
  {Khaitan}}, \bibinfo {author} {\bibfnamefont {A.}~\bibnamefont {Choudhary}},
  \ and\ \bibinfo {author} {\bibfnamefont {A.}~\bibnamefont {Agrawal}},\
  }\bibfield  {title} {\enquote {\bibinfo {title} {Deep convolutional neural
  networks with transfer learning for computer vision-based data-driven
  pavement distress detection},}\ }\href@noop {} {\bibfield  {journal}
  {\bibinfo  {journal} {Construction and building materials}\ }\textbf
  {\bibinfo {volume} {157}},\ \bibinfo {pages} {322--330} (\bibinfo {year}
  {2017})}\BibitemShut {NoStop}%
\bibitem [{\citenamefont {Ding}(2014)}]{ding2014data}%
  \BibitemOpen
  \bibfield  {author} {\bibinfo {author} {\bibfnamefont {S.~X.}\ \bibnamefont
  {Ding}},\ }\href@noop {} {\emph {\bibinfo {title} {Data-driven design of
  fault diagnosis and fault-tolerant control systems}}}\ (\bibinfo  {publisher}
  {Springer London},\ \bibinfo {year} {2014})\BibitemShut {NoStop}%
\bibitem [{\citenamefont {Hou}\ and\ \citenamefont
  {Wang}(2013)}]{hou2013model}%
  \BibitemOpen
  \bibfield  {author} {\bibinfo {author} {\bibfnamefont {Z.-S.}\ \bibnamefont
  {Hou}}\ and\ \bibinfo {author} {\bibfnamefont {Z.}~\bibnamefont {Wang}},\
  }\bibfield  {title} {\enquote {\bibinfo {title} {From model-based control to
  data-driven control: Survey, classification and perspective},}\ }\href@noop
  {} {\bibfield  {journal} {\bibinfo  {journal} {Information Sciences}\
  }\textbf {\bibinfo {volume} {235}},\ \bibinfo {pages} {3--35} (\bibinfo
  {year} {2013})}\BibitemShut {NoStop}%
\bibitem [{\citenamefont {Sarghini}, \citenamefont {De~Felice},\ and\
  \citenamefont {Santini}(2003)}]{sarghini2003neural}%
  \BibitemOpen
  \bibfield  {author} {\bibinfo {author} {\bibfnamefont {F.}~\bibnamefont
  {Sarghini}}, \bibinfo {author} {\bibfnamefont {G.}~\bibnamefont {De~Felice}},
  \ and\ \bibinfo {author} {\bibfnamefont {S.}~\bibnamefont {Santini}},\
  }\bibfield  {title} {\enquote {\bibinfo {title} {Neural networks based
  subgrid scale modeling in large eddy simulations},}\ }\href@noop {}
  {\bibfield  {journal} {\bibinfo  {journal} {Computers \& fluids}\ }\textbf
  {\bibinfo {volume} {32}},\ \bibinfo {pages} {97--108} (\bibinfo {year}
  {2003})}\BibitemShut {NoStop}%
\bibitem [{\citenamefont {Gamahara}\ and\ \citenamefont
  {Hattori}(2017)}]{gamahara2017searching}%
  \BibitemOpen
  \bibfield  {author} {\bibinfo {author} {\bibfnamefont {M.}~\bibnamefont
  {Gamahara}}\ and\ \bibinfo {author} {\bibfnamefont {Y.}~\bibnamefont
  {Hattori}},\ }\bibfield  {title} {\enquote {\bibinfo {title} {Searching for
  turbulence models by artificial neural network},}\ }\href@noop {} {\bibfield
  {journal} {\bibinfo  {journal} {Physical Review Fluids}\ }\textbf {\bibinfo
  {volume} {2}},\ \bibinfo {pages} {054604} (\bibinfo {year}
  {2017})}\BibitemShut {NoStop}%
\bibitem [{\citenamefont {Wu}, \citenamefont {Xiao},\ and\ \citenamefont
  {Paterson}(2018)}]{wu2018physics}%
  \BibitemOpen
  \bibfield  {author} {\bibinfo {author} {\bibfnamefont {J.-L.}\ \bibnamefont
  {Wu}}, \bibinfo {author} {\bibfnamefont {H.}~\bibnamefont {Xiao}}, \ and\
  \bibinfo {author} {\bibfnamefont {E.}~\bibnamefont {Paterson}},\ }\bibfield
  {title} {\enquote {\bibinfo {title} {Physics-informed machine learning
  approach for augmenting turbulence models: A comprehensive framework},}\
  }\href@noop {} {\bibfield  {journal} {\bibinfo  {journal} {Physical Review
  Fluids}\ }\textbf {\bibinfo {volume} {3}},\ \bibinfo {pages} {074602}
  (\bibinfo {year} {2018})}\BibitemShut {NoStop}%
\bibitem [{\citenamefont {Xie}\ \emph {et~al.}(2019)\citenamefont {Xie},
  \citenamefont {Wang}, \citenamefont {Li}, \citenamefont {Wan},\ and\
  \citenamefont {Chen}}]{xie2019artificial}%
  \BibitemOpen
  \bibfield  {author} {\bibinfo {author} {\bibfnamefont {C.}~\bibnamefont
  {Xie}}, \bibinfo {author} {\bibfnamefont {J.}~\bibnamefont {Wang}}, \bibinfo
  {author} {\bibfnamefont {H.}~\bibnamefont {Li}}, \bibinfo {author}
  {\bibfnamefont {M.}~\bibnamefont {Wan}}, \ and\ \bibinfo {author}
  {\bibfnamefont {S.}~\bibnamefont {Chen}},\ }\bibfield  {title} {\enquote
  {\bibinfo {title} {Artificial neural network mixed model for large eddy
  simulation of compressible isotropic turbulence},}\ }\href@noop {} {\bibfield
   {journal} {\bibinfo  {journal} {Physics of Fluids}\ }\textbf {\bibinfo
  {volume} {31}} (\bibinfo {year} {2019})}\BibitemShut {NoStop}%
\bibitem [{\citenamefont {Cai}\ \emph {et~al.}(2021)\citenamefont {Cai},
  \citenamefont {Wang}, \citenamefont {Fuest}, \citenamefont {Jeon},
  \citenamefont {Gray},\ and\ \citenamefont {Karniadakis}}]{cai2021flow}%
  \BibitemOpen
  \bibfield  {author} {\bibinfo {author} {\bibfnamefont {S.}~\bibnamefont
  {Cai}}, \bibinfo {author} {\bibfnamefont {Z.}~\bibnamefont {Wang}}, \bibinfo
  {author} {\bibfnamefont {F.}~\bibnamefont {Fuest}}, \bibinfo {author}
  {\bibfnamefont {Y.~J.}\ \bibnamefont {Jeon}}, \bibinfo {author}
  {\bibfnamefont {C.}~\bibnamefont {Gray}}, \ and\ \bibinfo {author}
  {\bibfnamefont {G.~E.}\ \bibnamefont {Karniadakis}},\ }\bibfield  {title}
  {\enquote {\bibinfo {title} {Flow over an espresso cup: inferring 3-d
  velocity and pressure fields from tomographic background oriented {S}chlieren
  via physics-informed neural networks},}\ }\href@noop {} {\bibfield  {journal}
  {\bibinfo  {journal} {Journal of Fluid Mechanics}\ }\textbf {\bibinfo
  {volume} {915}},\ \bibinfo {pages} {A102} (\bibinfo {year}
  {2021})}\BibitemShut {NoStop}%
\bibitem [{\citenamefont {Bae}\ and\ \citenamefont
  {Koumoutsakos}(2022)}]{bae2022scientific}%
  \BibitemOpen
  \bibfield  {author} {\bibinfo {author} {\bibfnamefont {H.~J.}\ \bibnamefont
  {Bae}}\ and\ \bibinfo {author} {\bibfnamefont {P.}~\bibnamefont
  {Koumoutsakos}},\ }\bibfield  {title} {\enquote {\bibinfo {title} {Scientific
  multi-agent reinforcement learning for wall-models of turbulent flows},}\
  }\href@noop {} {\bibfield  {journal} {\bibinfo  {journal} {Nature
  Communications}\ }\textbf {\bibinfo {volume} {13}},\ \bibinfo {pages} {1443}
  (\bibinfo {year} {2022})}\BibitemShut {NoStop}%
\bibitem [{\citenamefont {Yang}\ \emph {et~al.}(2019)\citenamefont {Yang},
  \citenamefont {Zafar}, \citenamefont {Wang},\ and\ \citenamefont
  {Xiao}}]{yang2019predictive}%
  \BibitemOpen
  \bibfield  {author} {\bibinfo {author} {\bibfnamefont {X.}~\bibnamefont
  {Yang}}, \bibinfo {author} {\bibfnamefont {S.}~\bibnamefont {Zafar}},
  \bibinfo {author} {\bibfnamefont {J.-X.}\ \bibnamefont {Wang}}, \ and\
  \bibinfo {author} {\bibfnamefont {H.}~\bibnamefont {Xiao}},\ }\bibfield
  {title} {\enquote {\bibinfo {title} {Predictive large-eddy-simulation wall
  modeling via physics-informed neural networks},}\ }\href@noop {} {\bibfield
  {journal} {\bibinfo  {journal} {Physical Review Fluids}\ }\textbf {\bibinfo
  {volume} {4}},\ \bibinfo {pages} {034602} (\bibinfo {year}
  {2019})}\BibitemShut {NoStop}%
\bibitem [{\citenamefont {Raissi}, \citenamefont {Perdikaris},\ and\
  \citenamefont {Karniadakis}(2019)}]{raissi2019physics}%
  \BibitemOpen
  \bibfield  {author} {\bibinfo {author} {\bibfnamefont {M.}~\bibnamefont
  {Raissi}}, \bibinfo {author} {\bibfnamefont {P.}~\bibnamefont {Perdikaris}},
  \ and\ \bibinfo {author} {\bibfnamefont {G.~E.}\ \bibnamefont
  {Karniadakis}},\ }\bibfield  {title} {\enquote {\bibinfo {title}
  {Physics-informed neural networks: A deep learning framework for solving
  forward and inverse problems involving nonlinear partial differential
  equations},}\ }\href@noop {} {\bibfield  {journal} {\bibinfo  {journal}
  {Journal of Computational physics}\ }\textbf {\bibinfo {volume} {378}},\
  \bibinfo {pages} {686--707} (\bibinfo {year} {2019})}\BibitemShut {NoStop}%
\bibitem [{\citenamefont {Wang}, \citenamefont {Wu},\ and\ \citenamefont
  {Xiao}(2017)}]{wang2017physics}%
  \BibitemOpen
  \bibfield  {author} {\bibinfo {author} {\bibfnamefont {J.-X.}\ \bibnamefont
  {Wang}}, \bibinfo {author} {\bibfnamefont {J.-L.}\ \bibnamefont {Wu}}, \ and\
  \bibinfo {author} {\bibfnamefont {H.}~\bibnamefont {Xiao}},\ }\bibfield
  {title} {\enquote {\bibinfo {title} {Physics-informed machine learning
  approach for reconstructing {R}eynolds stress modeling discrepancies based on
  dns data},}\ }\href@noop {} {\bibfield  {journal} {\bibinfo  {journal}
  {Physical Review Fluids}\ }\textbf {\bibinfo {volume} {2}},\ \bibinfo {pages}
  {034603} (\bibinfo {year} {2017})}\BibitemShut {NoStop}%
\bibitem [{\citenamefont {Davidson}(2022)}]{davidson2022using}%
  \BibitemOpen
  \bibfield  {author} {\bibinfo {author} {\bibfnamefont {L.}~\bibnamefont
  {Davidson}},\ }\bibfield  {title} {\enquote {\bibinfo {title} {Using machine
  learning for formulating new wall functions for {L}arge {E}ddy {S}imulation:
  A second attempt},}\ }\href@noop {} {\bibfield  {journal} {\bibinfo
  {journal} {Div. of Fluid Dynamics, Mechanics and Maritime Sciences, Chalmers
  University of Technology}\ } (\bibinfo {year} {2022})}\BibitemShut {NoStop}%
\bibitem [{\citenamefont {Xue}\ \emph {et~al.}(2020)\citenamefont {Xue},
  \citenamefont {Biferale}, \citenamefont {Sbragaglia},\ and\ \citenamefont
  {Toschi}}]{xue2020brownian}%
  \BibitemOpen
  \bibfield  {author} {\bibinfo {author} {\bibfnamefont {X.}~\bibnamefont
  {Xue}}, \bibinfo {author} {\bibfnamefont {L.}~\bibnamefont {Biferale}},
  \bibinfo {author} {\bibfnamefont {M.}~\bibnamefont {Sbragaglia}}, \ and\
  \bibinfo {author} {\bibfnamefont {F.}~\bibnamefont {Toschi}},\ }\bibfield
  {title} {\enquote {\bibinfo {title} {A lattice {B}oltzmann study on
  {B}rownian diffusion and friction of a particle in a confined multicomponent
  fluid},}\ }\href@noop {} {\bibfield  {journal} {\bibinfo  {journal} {Journal
  of Computational Science}\ }\textbf {\bibinfo {volume} {47}},\ \bibinfo
  {pages} {101113} (\bibinfo {year} {2020})}\BibitemShut {NoStop}%
\bibitem [{\citenamefont {Xue}\ \emph {et~al.}(2018)\citenamefont {Xue},
  \citenamefont {Sbragaglia}, \citenamefont {Biferale},\ and\ \citenamefont
  {Toschi}}]{xue2018effects}%
  \BibitemOpen
  \bibfield  {author} {\bibinfo {author} {\bibfnamefont {X.}~\bibnamefont
  {Xue}}, \bibinfo {author} {\bibfnamefont {M.}~\bibnamefont {Sbragaglia}},
  \bibinfo {author} {\bibfnamefont {L.}~\bibnamefont {Biferale}}, \ and\
  \bibinfo {author} {\bibfnamefont {F.}~\bibnamefont {Toschi}},\ }\bibfield
  {title} {\enquote {\bibinfo {title} {Effects of thermal fluctuations in the
  fragmentation of a nanoligament},}\ }\href@noop {} {\bibfield  {journal}
  {\bibinfo  {journal} {Physical Review E}\ }\textbf {\bibinfo {volume} {98}},\
  \bibinfo {pages} {012802} (\bibinfo {year} {2018})}\BibitemShut {NoStop}%
\bibitem [{\citenamefont {Xue}\ \emph {et~al.}(2021)\citenamefont {Xue},
  \citenamefont {Biferale}, \citenamefont {Sbragaglia},\ and\ \citenamefont
  {Toschi}}]{xue2021lattice}%
  \BibitemOpen
  \bibfield  {author} {\bibinfo {author} {\bibfnamefont {X.}~\bibnamefont
  {Xue}}, \bibinfo {author} {\bibfnamefont {L.}~\bibnamefont {Biferale}},
  \bibinfo {author} {\bibfnamefont {M.}~\bibnamefont {Sbragaglia}}, \ and\
  \bibinfo {author} {\bibfnamefont {F.}~\bibnamefont {Toschi}},\ }\bibfield
  {title} {\enquote {\bibinfo {title} {A lattice {B}oltzmann study of particle
  settling in a fluctuating multicomponent fluid under confinement},}\
  }\href@noop {} {\bibfield  {journal} {\bibinfo  {journal} {The European
  Physical Journal E}\ }\textbf {\bibinfo {volume} {44}},\ \bibinfo {pages}
  {1--10} (\bibinfo {year} {2021})}\BibitemShut {NoStop}%
\bibitem [{\citenamefont {Chiappini}\ \emph {et~al.}(2018)\citenamefont
  {Chiappini}, \citenamefont {Xue}, \citenamefont {Falcucci},\ and\
  \citenamefont {Sbragaglia}}]{chiappini2018ligament}%
  \BibitemOpen
  \bibfield  {author} {\bibinfo {author} {\bibfnamefont {D.}~\bibnamefont
  {Chiappini}}, \bibinfo {author} {\bibfnamefont {X.}~\bibnamefont {Xue}},
  \bibinfo {author} {\bibfnamefont {G.}~\bibnamefont {Falcucci}}, \ and\
  \bibinfo {author} {\bibfnamefont {M.}~\bibnamefont {Sbragaglia}},\ }\bibfield
   {title} {\enquote {\bibinfo {title} {Ligament break-up simulation through
  pseudo-potential lattice {B}oltzmann method},}\ }in\ \href@noop {} {\emph
  {\bibinfo {booktitle} {AIP Conference Proceedings}}},\ Vol.\ \bibinfo
  {volume} {1978}\ (\bibinfo {organization} {AIP Publishing},\ \bibinfo {year}
  {2018})\ p.\ \bibinfo {pages} {420003}\BibitemShut {NoStop}%
\bibitem [{\citenamefont {Chiappini}\ \emph {et~al.}(2019)\citenamefont
  {Chiappini}, \citenamefont {Sbragaglia}, \citenamefont {Xue},\ and\
  \citenamefont {Falcucci}}]{chiappini2019hydrodynamic}%
  \BibitemOpen
  \bibfield  {author} {\bibinfo {author} {\bibfnamefont {D.}~\bibnamefont
  {Chiappini}}, \bibinfo {author} {\bibfnamefont {M.}~\bibnamefont
  {Sbragaglia}}, \bibinfo {author} {\bibfnamefont {X.}~\bibnamefont {Xue}}, \
  and\ \bibinfo {author} {\bibfnamefont {G.}~\bibnamefont {Falcucci}},\
  }\bibfield  {title} {\enquote {\bibinfo {title} {Hydrodynamic behavior of the
  pseudopotential lattice {B}oltzmann method for interfacial flows},}\
  }\href@noop {} {\bibfield  {journal} {\bibinfo  {journal} {Physical Review
  E}\ }\textbf {\bibinfo {volume} {99}},\ \bibinfo {pages} {053305} (\bibinfo
  {year} {2019})}\BibitemShut {NoStop}%
\bibitem [{\citenamefont {Hou}\ \emph {et~al.}(1995)\citenamefont {Hou},
  \citenamefont {Sterling}, \citenamefont {Chen},\ and\ \citenamefont
  {Doolen}}]{hou1994lattice}%
  \BibitemOpen
  \bibfield  {author} {\bibinfo {author} {\bibfnamefont {S.}~\bibnamefont
  {Hou}}, \bibinfo {author} {\bibfnamefont {J.}~\bibnamefont {Sterling}},
  \bibinfo {author} {\bibfnamefont {S.}~\bibnamefont {Chen}}, \ and\ \bibinfo
  {author} {\bibfnamefont {G.}~\bibnamefont {Doolen}},\ }\bibfield  {title}
  {\enquote {\bibinfo {title} {A lattice {B}oltzmann subgrid model for high
  {R}eynolds number flows},}\ }\href@noop {} {\bibfield  {journal} {\bibinfo
  {journal} {Pattern formation and lattice gas automata}\ ,\ \bibinfo {pages}
  {151--166}} (\bibinfo {year} {1995})}\BibitemShut {NoStop}%
\bibitem [{\citenamefont {Toschi}\ and\ \citenamefont
  {Bodenschatz}(2009)}]{toschi2009lagrangian}%
  \BibitemOpen
  \bibfield  {author} {\bibinfo {author} {\bibfnamefont {F.}~\bibnamefont
  {Toschi}}\ and\ \bibinfo {author} {\bibfnamefont {E.}~\bibnamefont
  {Bodenschatz}},\ }\bibfield  {title} {\enquote {\bibinfo {title} {Lagrangian
  properties of particles in turbulence},}\ }\href@noop {} {\bibfield
  {journal} {\bibinfo  {journal} {Annual Review of Fluid Mechanics}\ }\textbf
  {\bibinfo {volume} {41}},\ \bibinfo {pages} {375--404} (\bibinfo {year}
  {2009})}\BibitemShut {NoStop}%
\bibitem [{\citenamefont {Karlin}, \citenamefont {Ferrante},\ and\
  \citenamefont {{\"O}ttinger}(1999)}]{karlin1999perfect}%
  \BibitemOpen
  \bibfield  {author} {\bibinfo {author} {\bibfnamefont {I.~V.}\ \bibnamefont
  {Karlin}}, \bibinfo {author} {\bibfnamefont {A.}~\bibnamefont {Ferrante}}, \
  and\ \bibinfo {author} {\bibfnamefont {H.~C.}\ \bibnamefont {{\"O}ttinger}},\
  }\bibfield  {title} {\enquote {\bibinfo {title} {Perfect entropy functions of
  the lattice {B}oltzmann method},}\ }\href@noop {} {\bibfield  {journal}
  {\bibinfo  {journal} {EPL (Europhysics Letters)}\ }\textbf {\bibinfo {volume}
  {47}},\ \bibinfo {pages} {182} (\bibinfo {year} {1999})}\BibitemShut
  {NoStop}%
\bibitem [{\citenamefont {Lallemand}\ and\ \citenamefont
  {Luo}(2000)}]{lallemand2000theory}%
  \BibitemOpen
  \bibfield  {author} {\bibinfo {author} {\bibfnamefont {P.}~\bibnamefont
  {Lallemand}}\ and\ \bibinfo {author} {\bibfnamefont {L.-S.}\ \bibnamefont
  {Luo}},\ }\bibfield  {title} {\enquote {\bibinfo {title} {Theory of the
  lattice {B}oltzmann method: Dispersion, dissipation, isotropy, {G}alilean
  invariance, and stability},}\ }\href@noop {} {\bibfield  {journal} {\bibinfo
  {journal} {Physical review E}\ }\textbf {\bibinfo {volume} {61}},\ \bibinfo
  {pages} {6546} (\bibinfo {year} {2000})}\BibitemShut {NoStop}%
\bibitem [{\citenamefont {Succi}(2001)}]{succi2001lattice}%
  \BibitemOpen
  \bibfield  {author} {\bibinfo {author} {\bibfnamefont {S.}~\bibnamefont
  {Succi}},\ }\href@noop {} {\emph {\bibinfo {title} {The {L}attice {B}oltzmann
  {E}quation for {F}luid {D}ynamics and {B}eyond}}}\ (\bibinfo  {publisher}
  {Oxford University Press},\ \bibinfo {year} {2001})\BibitemShut {NoStop}%
\bibitem [{\citenamefont {Kr{\"u}ger}\ \emph {et~al.}(2017)\citenamefont
  {Kr{\"u}ger}, \citenamefont {Kusumaatmaja}, \citenamefont {Kuzmin},
  \citenamefont {Shardt}, \citenamefont {Silva},\ and\ \citenamefont
  {Viggen}}]{kruger2017lattice}%
  \BibitemOpen
  \bibfield  {author} {\bibinfo {author} {\bibfnamefont {T.}~\bibnamefont
  {Kr{\"u}ger}}, \bibinfo {author} {\bibfnamefont {H.}~\bibnamefont
  {Kusumaatmaja}}, \bibinfo {author} {\bibfnamefont {A.}~\bibnamefont
  {Kuzmin}}, \bibinfo {author} {\bibfnamefont {O.}~\bibnamefont {Shardt}},
  \bibinfo {author} {\bibfnamefont {G.}~\bibnamefont {Silva}}, \ and\ \bibinfo
  {author} {\bibfnamefont {E.~M.}\ \bibnamefont {Viggen}},\ }\bibfield  {title}
  {\enquote {\bibinfo {title} {The lattice {B}oltzmann method},}\ }\href@noop
  {} {\bibfield  {journal} {\bibinfo  {journal} {Springer International
  Publishing}\ }\textbf {\bibinfo {volume} {10}},\ \bibinfo {pages} {978--3}
  (\bibinfo {year} {2017})}\BibitemShut {NoStop}%
\bibitem [{\citenamefont {Lallemand}\ \emph {et~al.}(2021)\citenamefont
  {Lallemand}, \citenamefont {Luo}, \citenamefont {Krafczyk},\ and\
  \citenamefont {Yong}}]{lallemand2021lattice}%
  \BibitemOpen
  \bibfield  {author} {\bibinfo {author} {\bibfnamefont {P.}~\bibnamefont
  {Lallemand}}, \bibinfo {author} {\bibfnamefont {L.-s.}\ \bibnamefont {Luo}},
  \bibinfo {author} {\bibfnamefont {M.}~\bibnamefont {Krafczyk}}, \ and\
  \bibinfo {author} {\bibfnamefont {W.-A.}\ \bibnamefont {Yong}},\ }\bibfield
  {title} {\enquote {\bibinfo {title} {The lattice {B}oltzmann method for
  nearly incompressible flows},}\ }\href@noop {} {\bibfield  {journal}
  {\bibinfo  {journal} {Journal of Computational Physics}\ }\textbf {\bibinfo
  {volume} {431}},\ \bibinfo {pages} {109713} (\bibinfo {year}
  {2021})}\BibitemShut {NoStop}%
\bibitem [{\citenamefont {Malaspinas}\ and\ \citenamefont
  {Sagaut}(2014)}]{malaspinas2014wall}%
  \BibitemOpen
  \bibfield  {author} {\bibinfo {author} {\bibfnamefont {O.}~\bibnamefont
  {Malaspinas}}\ and\ \bibinfo {author} {\bibfnamefont {P.}~\bibnamefont
  {Sagaut}},\ }\bibfield  {title} {\enquote {\bibinfo {title} {Wall model for
  large-eddy simulation based on the lattice {B}oltzmann method},}\ }\href@noop
  {} {\bibfield  {journal} {\bibinfo  {journal} {Journal of Computational
  Physics}\ }\textbf {\bibinfo {volume} {275}},\ \bibinfo {pages} {25--40}
  (\bibinfo {year} {2014})}\BibitemShut {NoStop}%
\bibitem [{\citenamefont {Latt}\ \emph {et~al.}(2008)\citenamefont {Latt},
  \citenamefont {Chopard}, \citenamefont {Malaspinas}, \citenamefont
  {Deville},\ and\ \citenamefont {Michler}}]{latt2008straight}%
  \BibitemOpen
  \bibfield  {author} {\bibinfo {author} {\bibfnamefont {J.}~\bibnamefont
  {Latt}}, \bibinfo {author} {\bibfnamefont {B.}~\bibnamefont {Chopard}},
  \bibinfo {author} {\bibfnamefont {O.}~\bibnamefont {Malaspinas}}, \bibinfo
  {author} {\bibfnamefont {M.}~\bibnamefont {Deville}}, \ and\ \bibinfo
  {author} {\bibfnamefont {A.}~\bibnamefont {Michler}},\ }\bibfield  {title}
  {\enquote {\bibinfo {title} {Straight velocity boundaries in the lattice
  {B}oltzmann method},}\ }\href@noop {} {\bibfield  {journal} {\bibinfo
  {journal} {Physical Review E}\ }\textbf {\bibinfo {volume} {77}},\ \bibinfo
  {pages} {056703} (\bibinfo {year} {2008})}\BibitemShut {NoStop}%
\bibitem [{\citenamefont {Musker}(1979)}]{musker1979explicit}%
  \BibitemOpen
  \bibfield  {author} {\bibinfo {author} {\bibfnamefont {A.}~\bibnamefont
  {Musker}},\ }\bibfield  {title} {\enquote {\bibinfo {title} {Explicit
  expression for the smooth wall velocity distribution in a turbulent boundary
  layer},}\ }\href@noop {} {\bibfield  {journal} {\bibinfo  {journal} {AIAA
  Journal}\ }\textbf {\bibinfo {volume} {17}},\ \bibinfo {pages} {655--657}
  (\bibinfo {year} {1979})}\BibitemShut {NoStop}%
\bibitem [{\citenamefont {Haussmann}\ \emph {et~al.}(2019)\citenamefont
  {Haussmann}, \citenamefont {Barreto}, \citenamefont {Kouyi}, \citenamefont
  {Rivi{\`e}re}, \citenamefont {Nirschl},\ and\ \citenamefont
  {Krause}}]{haussmann2019large}%
  \BibitemOpen
  \bibfield  {author} {\bibinfo {author} {\bibfnamefont {M.}~\bibnamefont
  {Haussmann}}, \bibinfo {author} {\bibfnamefont {A.~C.}\ \bibnamefont
  {Barreto}}, \bibinfo {author} {\bibfnamefont {G.~L.}\ \bibnamefont {Kouyi}},
  \bibinfo {author} {\bibfnamefont {N.}~\bibnamefont {Rivi{\`e}re}}, \bibinfo
  {author} {\bibfnamefont {H.}~\bibnamefont {Nirschl}}, \ and\ \bibinfo
  {author} {\bibfnamefont {M.~J.}\ \bibnamefont {Krause}},\ }\bibfield  {title}
  {\enquote {\bibinfo {title} {Large-eddy simulation coupled with wall models
  for turbulent channel flows at high {R}eynolds numbers with a lattice
  {B}oltzmann method: Application to {C}oriolis mass flowmeter},}\ }\href@noop
  {} {\bibfield  {journal} {\bibinfo  {journal} {Computers \& Mathematics with
  Applications}\ }\textbf {\bibinfo {volume} {78}},\ \bibinfo {pages}
  {3285--3302} (\bibinfo {year} {2019})}\BibitemShut {NoStop}%
\bibitem [{\citenamefont {Maeyama}\ \emph {et~al.}(2020)\citenamefont
  {Maeyama}, \citenamefont {Imamura}, \citenamefont {Osaka},\ and\
  \citenamefont {Kurimoto}}]{maeyama2020unsteady}%
  \BibitemOpen
  \bibfield  {author} {\bibinfo {author} {\bibfnamefont {H.}~\bibnamefont
  {Maeyama}}, \bibinfo {author} {\bibfnamefont {T.}~\bibnamefont {Imamura}},
  \bibinfo {author} {\bibfnamefont {J.}~\bibnamefont {Osaka}}, \ and\ \bibinfo
  {author} {\bibfnamefont {N.}~\bibnamefont {Kurimoto}},\ }\bibfield  {title}
  {\enquote {\bibinfo {title} {Unsteady turbulent flow simulation using lattice
  {B}oltzmann method with near-wall modeling},}\ }in\ \href@noop {} {\emph
  {\bibinfo {booktitle} {AIAA Aviation 2020 Forum}}}\ (\bibinfo {year} {2020})\
  p.\ \bibinfo {pages} {2565}\BibitemShut {NoStop}%
\bibitem [{\citenamefont {Wilhelm}, \citenamefont {Jacob},\ and\ \citenamefont
  {Sagaut}(2021)}]{wilhelm2021new}%
  \BibitemOpen
  \bibfield  {author} {\bibinfo {author} {\bibfnamefont {S.}~\bibnamefont
  {Wilhelm}}, \bibinfo {author} {\bibfnamefont {J.}~\bibnamefont {Jacob}}, \
  and\ \bibinfo {author} {\bibfnamefont {P.}~\bibnamefont {Sagaut}},\
  }\bibfield  {title} {\enquote {\bibinfo {title} {A new explicit algebraic
  wall model for {LES} of turbulent flows under adverse pressure gradient},}\
  }\href@noop {} {\bibfield  {journal} {\bibinfo  {journal} {Flow, Turbulence
  and Combustion}\ }\textbf {\bibinfo {volume} {106}},\ \bibinfo {pages}
  {1--35} (\bibinfo {year} {2021})}\BibitemShut {NoStop}%
\bibitem [{\citenamefont {Kuwata}\ and\ \citenamefont
  {Suga}(2021)}]{kuwata2021wall}%
  \BibitemOpen
  \bibfield  {author} {\bibinfo {author} {\bibfnamefont {Y.}~\bibnamefont
  {Kuwata}}\ and\ \bibinfo {author} {\bibfnamefont {K.}~\bibnamefont {Suga}},\
  }\bibfield  {title} {\enquote {\bibinfo {title} {Wall-modeled large eddy
  simulation of turbulent heat transfer by the lattice {B}oltzmann method},}\
  }\href@noop {} {\bibfield  {journal} {\bibinfo  {journal} {Journal of
  Computational Physics}\ }\textbf {\bibinfo {volume} {433}},\ \bibinfo {pages}
  {110186} (\bibinfo {year} {2021})}\BibitemShut {NoStop}%
\bibitem [{\citenamefont {Xue}, \citenamefont {Yao},\ and\ \citenamefont
  {Davidson}(2023)}]{xue2023wall}%
  \BibitemOpen
  \bibfield  {author} {\bibinfo {author} {\bibfnamefont {X.}~\bibnamefont
  {Xue}}, \bibinfo {author} {\bibfnamefont {H.-D.}\ \bibnamefont {Yao}}, \ and\
  \bibinfo {author} {\bibfnamefont {L.}~\bibnamefont {Davidson}},\ }\bibfield
  {title} {\enquote {\bibinfo {title} {Wall-modeled large-eddy simulation
  integrated with synthetic turbulence generator for multiple-relaxation-time
  lattice {B}oltzmann method},}\ }\href@noop {} {\bibfield  {journal} {\bibinfo
   {journal} {Physics of Fluids}\ }\textbf {\bibinfo {volume} {35}} (\bibinfo
  {year} {2023})}\BibitemShut {NoStop}%
\bibitem [{\citenamefont {Shur}\ \emph {et~al.}(2008)\citenamefont {Shur},
  \citenamefont {Spalart}, \citenamefont {Strelets},\ and\ \citenamefont
  {Travin}}]{shur2008hybrid}%
  \BibitemOpen
  \bibfield  {author} {\bibinfo {author} {\bibfnamefont {M.~L.}\ \bibnamefont
  {Shur}}, \bibinfo {author} {\bibfnamefont {P.~R.}\ \bibnamefont {Spalart}},
  \bibinfo {author} {\bibfnamefont {M.~K.}\ \bibnamefont {Strelets}}, \ and\
  \bibinfo {author} {\bibfnamefont {A.~K.}\ \bibnamefont {Travin}},\ }\bibfield
   {title} {\enquote {\bibinfo {title} {A hybrid {RANS-{LES}} approach with
  delayed-{DES} and wall-modelled {{LES}} capabilities},}\ }\href@noop {}
  {\bibfield  {journal} {\bibinfo  {journal} {International Journal of Heat and
  Fluid Flow}\ }\textbf {\bibinfo {volume} {29}},\ \bibinfo {pages}
  {1638--1649} (\bibinfo {year} {2008})}\BibitemShut {NoStop}%
\bibitem [{\citenamefont {Xue}, \citenamefont {Yao},\ and\ \citenamefont
  {Davidson}(2022)}]{xue2022synthetic}%
  \BibitemOpen
  \bibfield  {author} {\bibinfo {author} {\bibfnamefont {X.}~\bibnamefont
  {Xue}}, \bibinfo {author} {\bibfnamefont {H.-D.}\ \bibnamefont {Yao}}, \ and\
  \bibinfo {author} {\bibfnamefont {L.}~\bibnamefont {Davidson}},\ }\bibfield
  {title} {\enquote {\bibinfo {title} {Synthetic turbulence generator for
  lattice {B}oltzmann method at the interface between rans and {LES}},}\
  }\href@noop {} {\bibfield  {journal} {\bibinfo  {journal} {Physics of
  Fluids}\ }\textbf {\bibinfo {volume} {34}},\ \bibinfo {pages} {055118}
  (\bibinfo {year} {2022})}\BibitemShut {NoStop}%
\bibitem [{\citenamefont {Hoyas}\ and\ \citenamefont
  {Jim{\'e}nez}(2006)}]{hoyas2006scaling}%
  \BibitemOpen
  \bibfield  {author} {\bibinfo {author} {\bibfnamefont {S.}~\bibnamefont
  {Hoyas}}\ and\ \bibinfo {author} {\bibfnamefont {J.}~\bibnamefont
  {Jim{\'e}nez}},\ }\bibfield  {title} {\enquote {\bibinfo {title} {Scaling of
  the velocity fluctuations in turbulent channels up to {R}e $\tau$= 2003},}\
  }\href@noop {} {\bibfield  {journal} {\bibinfo  {journal} {Physics of
  Fluids}\ }\textbf {\bibinfo {volume} {18}},\ \bibinfo {pages} {011702}
  (\bibinfo {year} {2006})}\BibitemShut {NoStop}%
\bibitem [{\citenamefont {Lee}\ and\ \citenamefont
  {Moser}(2015)}]{lee2015direct}%
  \BibitemOpen
  \bibfield  {author} {\bibinfo {author} {\bibfnamefont {M.}~\bibnamefont
  {Lee}}\ and\ \bibinfo {author} {\bibfnamefont {R.~D.}\ \bibnamefont
  {Moser}},\ }\bibfield  {title} {\enquote {\bibinfo {title} {Direct numerical
  simulation of turbulent channel flow up to {R}e $\tau \approx$ 5200},}\
  }\href@noop {} {\bibfield  {journal} {\bibinfo  {journal} {Journal of Fluid
  Mechanics}\ }\textbf {\bibinfo {volume} {774}},\ \bibinfo {pages} {395--415}
  (\bibinfo {year} {2015})}\BibitemShut {NoStop}%
\bibitem [{\citenamefont {Pasquali}, \citenamefont {Geier},\ and\ \citenamefont
  {Krafczyk}(2020)}]{pasquali2020near}%
  \BibitemOpen
  \bibfield  {author} {\bibinfo {author} {\bibfnamefont {A.}~\bibnamefont
  {Pasquali}}, \bibinfo {author} {\bibfnamefont {M.}~\bibnamefont {Geier}}, \
  and\ \bibinfo {author} {\bibfnamefont {M.}~\bibnamefont {Krafczyk}},\
  }\bibfield  {title} {\enquote {\bibinfo {title} {Near-wall treatment for the
  simulation of turbulent flow by the cumulant lattice {B}oltzmann method},}\
  }\href@noop {} {\bibfield  {journal} {\bibinfo  {journal} {Computers \&
  Mathematics with Applications}\ }\textbf {\bibinfo {volume} {79}},\ \bibinfo
  {pages} {195--212} (\bibinfo {year} {2020})}\BibitemShut {NoStop}%
\bibitem [{\citenamefont {d'Humieres}(2002)}]{d2002multiple}%
  \BibitemOpen
  \bibfield  {author} {\bibinfo {author} {\bibfnamefont {D.}~\bibnamefont
  {d'Humieres}},\ }\bibfield  {title} {\enquote {\bibinfo {title}
  {Multiple--relaxation--time lattice {B}oltzmann models in three
  dimensions},}\ }\href@noop {} {\bibfield  {journal} {\bibinfo  {journal}
  {Philosophical Transactions of the Royal Society of London. Series A:
  Mathematical, Physical and Engineering Sciences}\ }\textbf {\bibinfo {volume}
  {360}},\ \bibinfo {pages} {437--451} (\bibinfo {year} {2002})}\BibitemShut
  {NoStop}%
\bibitem [{\citenamefont {Guo}, \citenamefont {Zheng},\ and\ \citenamefont
  {Shi}(2002)}]{guo2002discrete}%
  \BibitemOpen
  \bibfield  {author} {\bibinfo {author} {\bibfnamefont {Z.}~\bibnamefont
  {Guo}}, \bibinfo {author} {\bibfnamefont {C.}~\bibnamefont {Zheng}}, \ and\
  \bibinfo {author} {\bibfnamefont {B.}~\bibnamefont {Shi}},\ }\bibfield
  {title} {\enquote {\bibinfo {title} {Discrete lattice effects on the forcing
  term in the lattice {B}oltzmann method},}\ }\href@noop {} {\bibfield
  {journal} {\bibinfo  {journal} {Physical review E}\ }\textbf {\bibinfo
  {volume} {65}},\ \bibinfo {pages} {046308} (\bibinfo {year}
  {2002})}\BibitemShut {NoStop}%
\bibitem [{\citenamefont {Smagorinsky}(1963)}]{smagorinsky1963general}%
  \BibitemOpen
  \bibfield  {author} {\bibinfo {author} {\bibfnamefont {J.}~\bibnamefont
  {Smagorinsky}},\ }\bibfield  {title} {\enquote {\bibinfo {title} {General
  circulation experiments with the primitive equations: I. the basic
  experiment},}\ }\href@noop {} {\bibfield  {journal} {\bibinfo  {journal}
  {Monthly Weather Review}\ }\textbf {\bibinfo {volume} {91}},\ \bibinfo
  {pages} {99--164} (\bibinfo {year} {1963})}\BibitemShut {NoStop}%
\bibitem [{\citenamefont {Koda}\ and\ \citenamefont
  {Lien}(2015)}]{koda2015lattice}%
  \BibitemOpen
  \bibfield  {author} {\bibinfo {author} {\bibfnamefont {Y.}~\bibnamefont
  {Koda}}\ and\ \bibinfo {author} {\bibfnamefont {F.-S.}\ \bibnamefont
  {Lien}},\ }\bibfield  {title} {\enquote {\bibinfo {title} {The lattice
  {B}oltzmann method implemented on the gpu to simulate the turbulent flow over
  a square cylinder confined in a channel},}\ }\href@noop {} {\bibfield
  {journal} {\bibinfo  {journal} {Flow, Turbulence and Combustion}\ }\textbf
  {\bibinfo {volume} {94}},\ \bibinfo {pages} {495--512} (\bibinfo {year}
  {2015})}\BibitemShut {NoStop}%
\bibitem [{\citenamefont {Abe}, \citenamefont {Kondoh},\ and\ \citenamefont
  {Nagano}(1994)}]{abe1994new}%
  \BibitemOpen
  \bibfield  {author} {\bibinfo {author} {\bibfnamefont {K.}~\bibnamefont
  {Abe}}, \bibinfo {author} {\bibfnamefont {T.}~\bibnamefont {Kondoh}}, \ and\
  \bibinfo {author} {\bibfnamefont {Y.}~\bibnamefont {Nagano}},\ }\bibfield
  {title} {\enquote {\bibinfo {title} {{A new turbulence model for predicting
  fluid flow and heat transfer in separating and reattaching flows—I. Flow
  field calculations}},}\ }\href@noop {} {\bibfield  {journal} {\bibinfo
  {journal} {International journal of heat and mass transfer}\ }\textbf
  {\bibinfo {volume} {37}},\ \bibinfo {pages} {139--151} (\bibinfo {year}
  {1994})}\BibitemShut {NoStop}%
\bibitem [{\citenamefont {Davidson}(2021)}]{davidson2021pycalc}%
  \BibitemOpen
  \bibfield  {author} {\bibinfo {author} {\bibfnamefont {L.}~\bibnamefont
  {Davidson}},\ }\bibfield  {title} {\enquote {\bibinfo {title} {{pyCALC-LES: A
  python code for DNS, LES and hybrid LES-RANS}},}\ }\href@noop {} {\bibfield
  {journal} {\bibinfo  {journal} {Chalmers University of Technology,
  Gothenburg}\ } (\bibinfo {year} {2021})}\BibitemShut {NoStop}%
\bibitem [{\citenamefont {Olson}\ and\ \citenamefont
  {Schroder}(2018)}]{olson2018pyamg}%
  \BibitemOpen
  \bibfield  {author} {\bibinfo {author} {\bibfnamefont {L.}~\bibnamefont
  {Olson}}\ and\ \bibinfo {author} {\bibfnamefont {J.}~\bibnamefont
  {Schroder}},\ }\bibfield  {title} {\enquote {\bibinfo {title} {{PyAMG:
  Algebraic multigrid solvers in Python v4. 0}},}\ }\href@noop {} {\bibfield
  {journal} {\bibinfo  {journal} {URL https://github. com/pyamg/pyamg.
  Release}\ }\textbf {\bibinfo {volume} {4}} (\bibinfo {year}
  {2018})}\BibitemShut {NoStop}%
\end{thebibliography}%

\section{Acknowledgements}
We acknowledge funding support from European Commission CompBioMed Centre of Excellence (Grant No. 675451 and 823712). Support from the UK Engineering and Physical Sciences Research Council under the projects ``UK Consortium on Mesoscale Engineering Sciences (UKCOMES)" (Grant No.
EP/R029598/1) and ``Software Environment for Actionable and VVUQ-evaluated Exascale Applications (SEAVEA)" (Grant No. EP/W007711/1) is gratefully acknowledged. We kindly acknowledge the funding from Chalmers Transport Area of Advance.


\section{Declaration of interests}
The authors declare that they have no known competing financial interests or personal relationships that could have appeared to influence the work reported in this paper.
\section{Material availability}
The training code and data set for sparse data and dense data can be found on University College London Centre for Computational Science GitHub page: \url{https://github.com/UCL-CCS/PINN-WM-LBM}.

\end{document}